\documentclass[%
 reprint,
 amsmath,amssymb,
 aps,
]{revtex4-1}

\usepackage{textcomp}
\usepackage{gensymb}
\usepackage{graphicx}% Include figure files
\usepackage{dcolumn}% Align table columns on decimal point
\usepackage{bm}% bold math
\usepackage{hyperref}% add hypertext capabilities
\PassOptionsToPackage{unicode}{hyperref}
\PassOptionsToPackage{naturalnames}{hyperref}
\usepackage[version=3]{mhchem}
\usepackage{xcolor}
\usepackage[caption=false]{subfig}
\renewcommand{\figurename}{\textbf{Fig.}}
\usepackage{hyperref, color, soul}
\definecolor{citeColor}{rgb}{0,0,0.7}
\definecolor{linkColor}{rgb}{0,0.6,0.2}
\hypersetup{pdfborder={0 0 0},colorlinks=true,urlcolor=citeColor,citecolor=citeColor,linkcolor=linkColor}
\emergencystretch=\maxdimen
\setcitestyle{super}
\usepackage{titlesec}
\titleformat{\section}{\centering\normalsize\bfseries}{}{}{}
\usepackage[T1]{fontenc}

\usepackage{tabularx}  % tables
\usepackage{multirow}
\usepackage{booktabs}
\usepackage{xcolor}

\begin{document}

\title{Ultrafast optical melting of trimer superstructure in layered 1T'-\ce{TaTe2}}

\author{Khalid M. Siddiqui\textit{$^{1}$}}\altaffiliation{These authors contributed equally to this work.}
\author{Daniel B. Durham\textit{$^{2,3}$}}\altaffiliation{These authors contributed equally to this work.}
\author{Frederick Cropp\textit{$^{4,5}$}}
\author{Colin Ophus\textit{$^{6}$}}
\author{Sangeeta Rajpurohit\textit{$^{6}$}}
\author{Yanglin Zhu\textit{$^{7}$}}
\author{Johan D. Carlstr\"{o}m\textit{$^{6}$}}
\author{Camille Stavrakas\textit{$^{6}$}}
\author{Zhiqiang Mao\textit{$^{7}$}}
\author{Archana Raja\textit{$^{6}$}}
\author{Pietro Musumeci\textit{$^{4}$}}
\author{Liang Z. Tan\textit{$^{6}$}}
\author{Andrew M. Minor\textit{$^{2,3}$}}
\author{Daniele Filippetto\textit{$^{5}$}}\altaffiliation{email: dfilippetto@lbl.gov; kaindl@asu.edu.}
\author{Robert A. Kaindl\textit{$^{1,8}$}}\altaffiliation{email: dfilippetto@lbl.gov; kaindl@asu.edu.}
\affiliation{\textit{$^{1}$}Materials Sciences Division, Lawrence Berkeley National Laboratory, Berkeley, California 94720, USA}
\affiliation{\textit{$^{2}$}National Center for Electron Microscopy, Molecular Foundry, Lawrence Berkeley National Laboratory,
Berkeley, California 94720, USA}
\affiliation{\textit{$^{3}$}Department of Materials Science and Engineering, University of California at Berkeley, Berkeley, California 94720, USA}
\affiliation{\textit{$^{4}$}Department of Physics and Astronomy, University of California Los Angeles, Los Angeles, California 90095, USA}
\affiliation{\textit{$^{5}$}Accelerator Technology and Applied Physics Division, Lawrence Berkeley National Laboratory, Berkeley, California 94720, USA}
\affiliation{\textit{$^{6}$}Molecular Foundry, Lawrence Berkeley National Laboratory, Berkeley, California 94720, USA}
\affiliation{\textit{$^{7}$}Department of Physics, The Pennsylvania State University, University Park, PA 16802, USA}
\affiliation{\textit{$^{8}$}Department of Physics, Arizona State University, Tempe, Arizona 85287, USA}

\begin{abstract}
Quasi-two-dimensional transition-metal dichalcogenides are a key platform for exploring emergent nanoscale phenomena arising from complex interactions. Access to the underlying degrees-of-freedom on their natural time scales motivates the use of advanced ultrafast probes sensitive to self-organised atomic-scale patterns. Here, we report the first ultrafast investigation of \ce{TaTe2}, which exhibits unique charge and lattice trimer order characterised by a transition upon cooling from stripe-like chains into a $(3 \times 3)$ superstructure of trimer clusters. Utilising MeV-scale ultrafast electron diffraction, we capture the photo-induced \ce{TaTe2} structural dynamics -- exposing a rapid $\approx\!1.4$ ps melting of its low-temperature ordered state followed by recovery via thermalisation into a hot cluster superstructure. Density-functional calculations indicate that the initial quench is triggered by intra-trimer Ta charge transfer which destabilises the clusters, unlike melting of charge density waves in other \ce{TaX2} compounds. Our work paves the way for further exploration and ultimately rapid optical and electronic manipulation of trimer superstructures.
\end{abstract}

\maketitle

\noindent Harnessing emergent orders in quantum materials has the potential to revolutionise energy and information technologies \cite{Tok17}. Complex interactions between lattice, electron, and spin degrees of freedom in these systems can give rise to emergent physics such as unconventional superconductivity \cite{Kei15}, topological protection \cite{Has10,Arm18}, or charge density wave (CDW) order and tailored interactions in two-dimensional (2D) materials \cite{Nov16}. Traditionally, control of materials has been achieved by adiabatic tuning of external parameters. Alternatively, ultrashort light pulses can be employed to perturb and transform states in quantum materials on femtosecond time scales \cite{Zha14,Bas17}. The quest to probe and control electronic and lattice structural dynamics in solids has driven the utilisation of advanced ultrafast spectroscopies, including ultrafast X-ray and electron diffraction \cite{Cav01,Lee12,Tri13,Otto_2018,Sie19,Zon19}, multi-terahertz fields \cite{Rin07,Por14,Cos17}, and time-resolved photoemission \cite{Per06,Roh11,Sob20}.  

\begin{figure*}
\centering
\includegraphics[width=15cm]{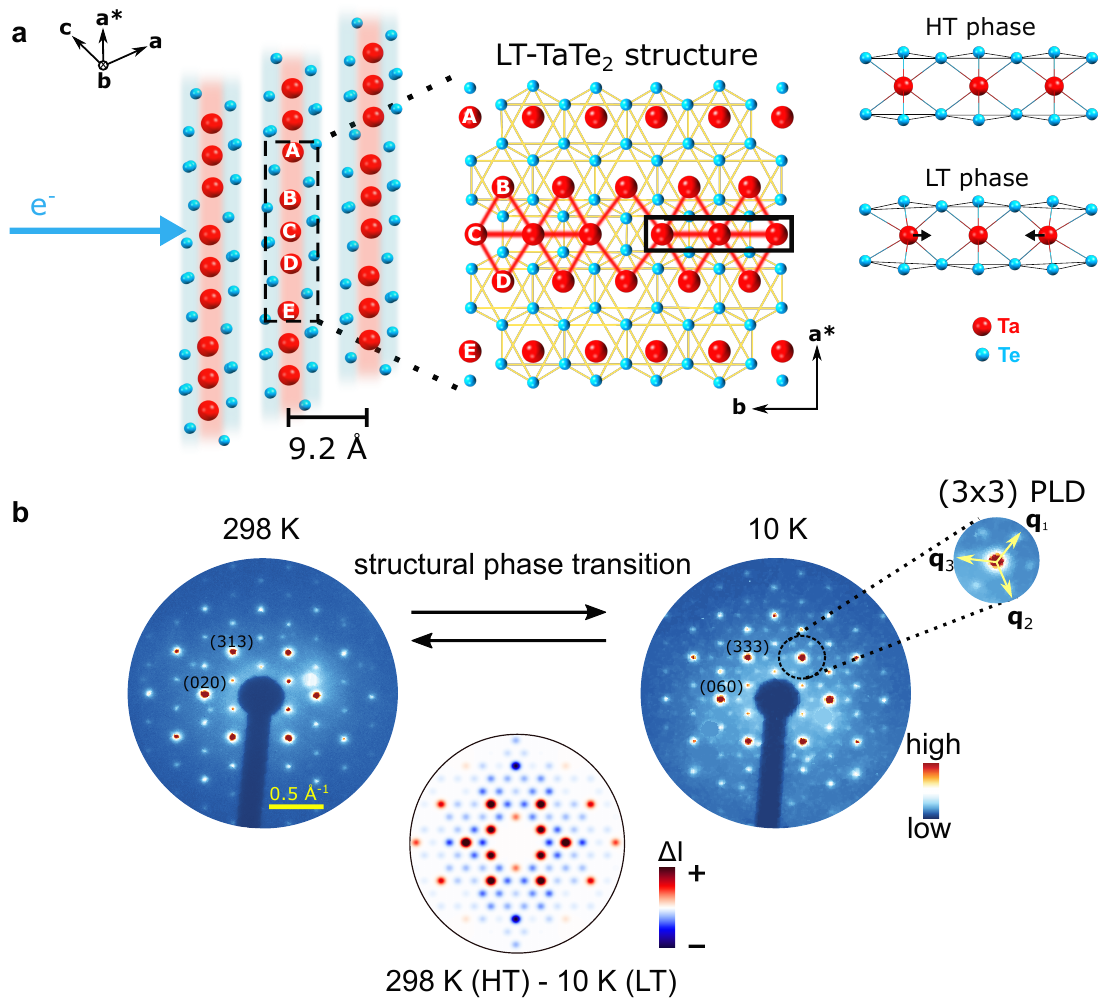}
\caption{\textbf{Crystal structure and electron diffraction patterns at thermal equilibiria. a} Layered crystal structure of the low temperature phase of 1T'-\ce{TaTe2}. The electron beam is incident perpendicular to Ta and Te planes in the UED experiments. The a, b, and c lattice vectors shown are those of the monoclinic unit cell. The projection perpendicular to the dashed region is indicated. Black box: Ta trimer clusters forming along the b-axis in the LT phase. Representations of LT and HT phases showing enhanced distortions in the LT phase are presented in the top right. Small black arrows denote the movement of atoms towards the central Ta atom. \textbf{b} Static electron diffraction patterns of 1T'-\ce{TaTe2} obtained by 0.75 MeV electron pulses at 298~K (HT) and 10~K (LT) along the $[\bar{1}01]$ zone axis. Note the different indices for LT pattern due to tripling along \textit{b}-axis. The inset shows the (3x3) PLD satellite peaks  that arise in the LT phase. A few additional peaks are present due to diffraction from the Si support frame (see Supplementary Note 2 for sample details). The symmetrised difference pattern between HT and LT phases is shown below the static patterns, with contributions from the Si frame and \ce{Si3N4} membrane removed.}
\label{10KStaticDiff}
\end{figure*}

Tantalum dichalcogenides (\ce{TaX2}, X = S, Se, Te) represent a class of materials that are well-matched for this pursuit and have gained increasing attention as quasi-2D systems with enhanced Coulomb and electron-lattice interactions \cite{Sip08,Cho16,Miller2018}. These compounds exhibit rich phase diagrams including semi-metallic, charge-ordered, and superconducting behaviours. Multiple CDW phases are observed in 1T-\ce{TaS2} which has spawned numerous ultrafast studies to clarify the formation mechanisms and phase competition \cite{Eichberger_2010,Haupt2016,Vogelgesang_2017}. Moreover, ultrafast driving exposed novel metastable phases in these systems, resulting in a new paradigm of hidden states \cite{Sto14,CYRuan2015TaS2}. Members of the $\text{1T-TaSe}_{2-x}\text{Te}_{x}$ family exhibit varying polytypes, CDW ground states, as well as superconductivity depending on $x$ \cite{Luo_2015}. Ultrafast optical melting and switching between CDW phases have so far been demonstrated in \ce{TaSe2} and in $\text{TaSe}_{2-x}\text{Te}_{x}$ alloys, accessing a range of dynamical pathways and timescales \cite{Sun2015, Linlin2017, li2019ultrafast}.

Curiously eluding ultrafast investigation thus far is \ce{TaTe2}. This compound exhibits markedly different properties with respect to the other Ta dichalcogenides, attributed to weaker electronegativity of Te with respect to Ta leading to a strong propensity for charge transfer and metal-metal bonding \cite{Wil69,Sorgel2006,Doublet_2000}. \ce{TaTe2} exhibits stronger electron-phonon coupling, higher charge-order binding energy, and larger lattice distortions than \ce{TaS2} and \ce{TaSe2} \cite{Miller2018,Gao2018}.  The room temperature  distorted monoclinic 1T' crystal structure is characterised by an intra-layer (${3\times1}$) linear stripe-like order composed of double zigzag Ta trimer chains. A structural transition into a phase with ${(3\times3)}$ order occurs at $T_{\rm PT} = 174$~K, with Ta atoms forming trimer clusters along the linear chains with commensurate CDW-like order \cite{Sorgel2006,Che18,Bag18,Wan20}. Unlike other \ce{TaX2} CDW systems \cite{Svetin_2017,LeBlanc_2010}, the low temperature phase ordering of \ce{TaTe2} exhibits metallic behaviour, with enhanced conductivity and magnetic susceptibility \cite{Chen_2017}. However, the ultrafast response of this compound to optical driving remains unknown. This motivates the use of advanced structural probes to follow the evolution of distortions and periodic order, as a measure of underlying interactions.

We report the first ultrafast study of \ce{TaTe2}, demonstrating a rapid picosecond melting of its trimer-cluster lattice superstructure in the low-temperature phase. Ultrafast electron diffraction (UED) with relativistic electron bunches is applied using the High Repetition-rate Electron Scattering (HiRES) beamline \cite{HiRESsims, ji_ultrafast_2019,Siddiqui_2020} to probe the time evolution of lattice order after intense near-infrared (near-IR) excitation. We observe photo-induced melting of the low-temperature order on a $\approx\!1.4$~ps time scale, indicative of fast switching, followed by recovery into a hot ${(3\times3)}$ trimer phase. Insight into the nature of trimer cluster melting is obtained via density functional calculations, which indicate an initial quench driven by charge transfer transitions from bonding to non-bonding states of the Ta trimer -- suggesting pathways for a photo-induced transition that is unique among the family of \ce{TaX2} materials. This work establishes \ce{TaTe2} as a promising material for optical control, motivating examination of concomitant electronic dynamics for device applications.

\begin{figure*}
    \centering
    \includegraphics[width=17cm]{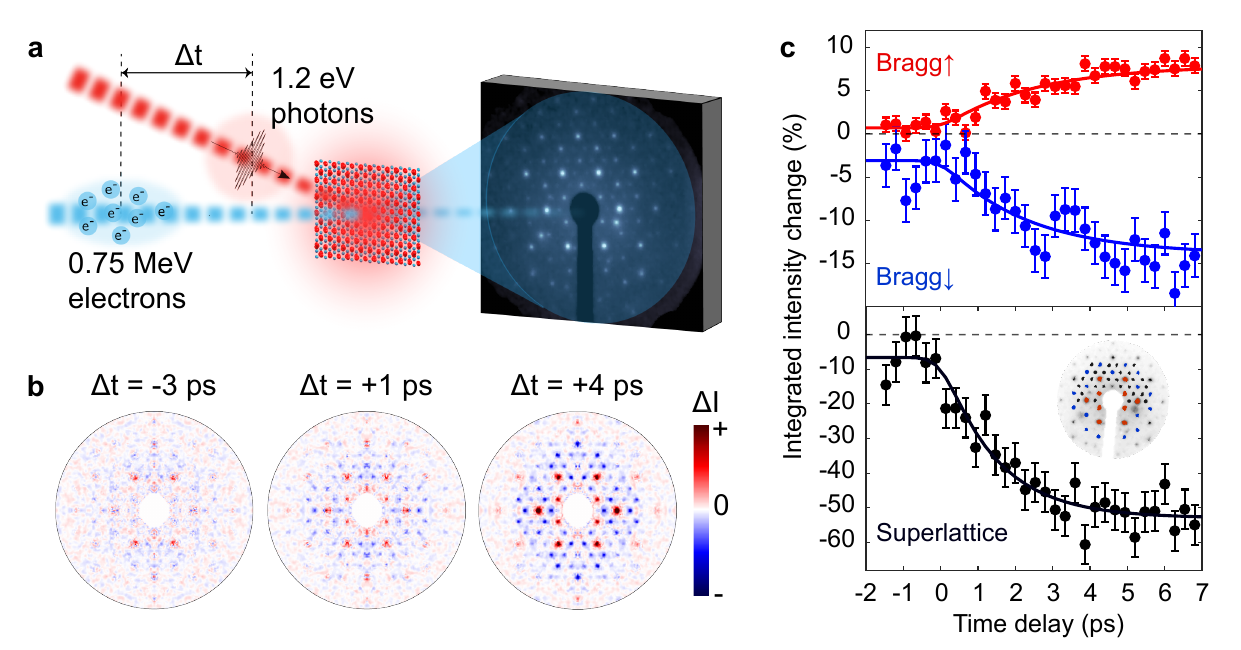}
\caption{\textbf{Ultrafast electron diffraction of \ce{TaTe2}. a} Illustration of the optical pump, ultrafast electron diffraction (UED) probe experiment. \textbf{ b} Photo-induced changes in the diffraction patterns, for selected time delays. The patterns have been symmetrised for visualisation purposes only. \textbf{c} Temporal evolution of the signals from the main lattice Bragg peaks, separately for the increasing and decreasing subsets, and of the LT superlattice peaks. Solid lines: fits with an exponential function convolved with the time resolution of 0.75 ps, up to 7 ps delay time. The corresponding time constants are $\tau_{\text{PLD}} = 1.44 \pm 0.27 $ ps, $\tau_{\text{Bragg}\uparrow} = 2.44 \pm 0.81$ ps and $\tau_{\text{Bragg}\downarrow} = 2.25 \pm 1.34$ ps. Error bars in the data indicate standard error calculated using the distribution of laser-off signals compared to the mean laser-off signal over the course of the measurement.}
    \label{relChangePlots}
\end{figure*}

\section*{Results}

\noindent {\bf Crystal structure and signatures of structural phase transition.} Figure~\ref{10KStaticDiff}a illustrates the crystal structure of \ce{TaTe2} in its low-temperature (LT) phase. Triple-layer sheets of covalently bonded Ta and Te atoms are separated by weaker van der Waals forces along the stacking direction. Prominent structural elements in this material are Ta trimers \textemdash sets of three adjacent Ta atoms in a row that cluster together via enhanced Ta-Ta bonding. Already at room temperature, Ta atoms are ordered in-plane into trimers assembled into double zigzag chains along the $b$-axis, which breaks hexagonal symmetry and forms a three-layer stacking sequence. In the LT phase, additional ordering emerges in the chains along the $b$-axis in the form of trimer clusters. This ${(3\times3)}$ lattice superstructure represents a distorted 1T' polytype with C2/m space group symmetry and a monoclinic unit cell \cite{Vernes_1998}. In this configuration, each Ta atom in the unit cell is coordinated to six Te atoms in a periodically distorted octahedral arrangement. 

Our DFT calculations of relaxed structures of \ce{TaTe2} confirm the distortions due to an atomic ordering of Ta atoms attributed to metal-metal bonding, in agreement with the structures previously determined by X-ray diffraction \cite{Sorgel2006}. Moreover, our calculations also reveal enhanced Ta-Ta bonds along the $b$-axis in the LT phase as highlighted in Fig.~\ref{10KStaticDiff}a, resulting in the tripling along the $b$-axis corresponding to an overall ${(3\times3)}$  superstructure (cf. Supplementary Note 1 for additional details).

Figure~\ref{10KStaticDiff}b shows equilibrium diffraction patterns of a 1T'-\ce{TaTe2} flake which we measured with the 0.75 MeV electron bunches at HiRES, comparing the high-temperature (HT) phase at 298~K and LT phase at 10~K. As illustrated in Fig.~\ref{10KStaticDiff}a, the electron beam impinges along the $[\bar{1}01]$ zone axis, i.e. perpendicular to the Ta and Te layers. The measured diffraction patterns exhibit a large number of Bragg spots reaching up to high momentum transfer, demonstrating both a high sample crystallinity and a large scattering range afforded by the relativistic beam energy. More details about the sample and its preparation, including transport measurements, are given in the Methods and in Supplementary Note 2.

At 298~K, we observe 2-fold symmetry exemplified, for instance, by differences of the (020) and (313) Bragg peaks in intensity and their relative distance from the centre. This is consistent with the (${3\times1}$) periodicity and the monoclinic crystal structure.

The pattern at 10 K in Fig.~\ref{10KStaticDiff}b reveals the appearance of new satellite peaks surrounding the main lattice peaks as a result of the emergent ${(3\times3)}$ periodic lattice distortion (PLD), in concordance with Ta trimer cluster formation and the associated unit cell tripling \cite{Sorgel2006,Che18, Bag18,Wan20}. Their observation also demonstrates that the transverse coherence length of the electron source is sufficient to track the dynamics of the LT superstructure in 1T'-\ce{TaTe2}. Analogous satellite peaks in the HT phase are $\approx\!1000$~times weaker than the main Bragg peaks and are not observed in these measurements (see Supplementary Note 2).

To determine the signature in the electron diffraction patterns attributed to the  structural phase transition, we calculate the difference between HT and LT patterns following normalisation by the total electron intensity. The resulting changes are shown at the bottom of Fig.~\ref{10KStaticDiff}b.  While all superlattice satellites associated with the ${(3\times3)}$ trimer superstructure are suppressed, the main Bragg peaks exhibit a mixture of positive and negative intensity changes. This complex response deviates from observations in \ce{TaS2} and \ce{TaSe2} where all primary Bragg peaks increased in intensity, opposite to the suppression of the PLD satellites \cite{Eichberger_2010,Erasmus2012}. We attribute this positive-negative intensity change signature to the symmetry of the superstructure formation within the distorted monoclinic unit cell, leading to mixed structure factor changes for different diffraction orders as supported by our simulations (see Supplementary Note 3).\cite{Kirkland_2010,Ophus_2017}

\vspace{3 mm}

\noindent {\bf Ultrafast optical melting of trimer clusters.} We utilise the HiRES beamline for ultrafast electron diffraction. The sample was first cooled into the low-temperature ordered phase at 10~K, and then photo-excited with near-IR femtosecond pulses (1030 nm wavelength). Time-delayed electron pulses at 0.75~MeV are used as structural probe, as illustrated in Fig.~\ref{relChangePlots}a. Recorded diffraction patterns with and without excitation provide signatures of photo-induced changes for each pump-probe time delay.

Difference maps of the diffraction intensity at selected time delays $\Delta t$ are shown in Fig.~\ref{relChangePlots}b for a pump fluence of 2.3~mJ cm$^{-2}$, indicating structural changes on a picosecond timescale. For clearer visualisation, these maps are symmetrised by averaging signals across the vertical and horizontal mirror plane symmetries (the process is described in Supplementary Note 4). At early times the signals exhibit a characteristic pattern of changes, including a decrease of the ${(3\times3)}$ PLD satellites and a mixture of increased and decreased intensities of the main lattice Bragg peaks (see diffraction pattern at $\Delta t = 4$ ps in Fig.~\ref{relChangePlots}b). This pattern strongly resembles that of the HT-LT phase transition obtained from the equilibrium data in Fig.~\ref{10KStaticDiff}a. This indicates that the intense near-IR excitation induces a melting of the low-temperature trimer clusters and an ultrafast phase transition to the (${3\times1}$) ordered HT state in \ce{TaTe2}.

In order to track the structural kinetics, we fit all peaks for each time-delayed UED pattern, summing the photo-induced changes of specific subsets for optimal signal-to-noise (peak fitting procedure is provided in Supplementary Note 5). Figure~\ref{relChangePlots}c plots the dynamics of the set of lattice Bragg peaks exhibiting an intensity increase (Bragg$\uparrow$) or decrease (Bragg$\downarrow$) in the pattern, as well as the changes of  the superlattice satellites. Normalised to their intensities before excitation, the PLD satellites undergo $\approx\!55$\% suppression with a time constant of $\tau_{\text{PLD}}~\approx\!1.4$~ps (details of the fitting are provided in Supplementary Note 6). This time constant provides a measure of the (${3\times3}$) trimer superstructure melting time in \ce{TaTe2} in our experiments. We note this is likely preceded by a faster electronic melting time which we cannot access here, but could be a subject of future spectroscopic investigations \cite{Per06}.

Alongside the PLD suppression, the primary lattice Bragg peaks also exhibit strong changes with slower dynamics corresponding to time constants of  ${\tau_{\text{Bragg$\uparrow$}}~\approx~\!2.4}$~ps and ${\tau_{\text{ Bragg$\downarrow$}}~\approx~\!2.3}$~ps. The underlying diffraction orders show comparable dynamics (see Supplementary Note~7). Moreover, an oscillation seems to appear in the Bragg$\downarrow$ trace, with a period of $\approx\!2$~ps. While this may be linked to excitation of coherent phonons, the $\approx\!0.5$~THz frequency does not match vibrational modes identified by theory (cf. Supplementary Note 1). The lack of a similar feature on the Bragg$\uparrow$ curve does not rule out a coherent phonon origin. However, the error bars in the Bragg$\downarrow$ trace are larger due to its weaker constituent high-$q$ peaks (cf. Supplementary Note 7) and the fluctuation is comparable to measurement error. While beyond the scope of our present work, future investigations are warranted to clarify the presence of coherent lattice motion. We also note that at negative time delays, a few-percent intensity reduction is observed in both the superlattice and Bragg$\downarrow$ peaks which we attribute to residual heating accumulated over several laser pulses.

\begin{figure}
    \includegraphics[width=8.5cm]{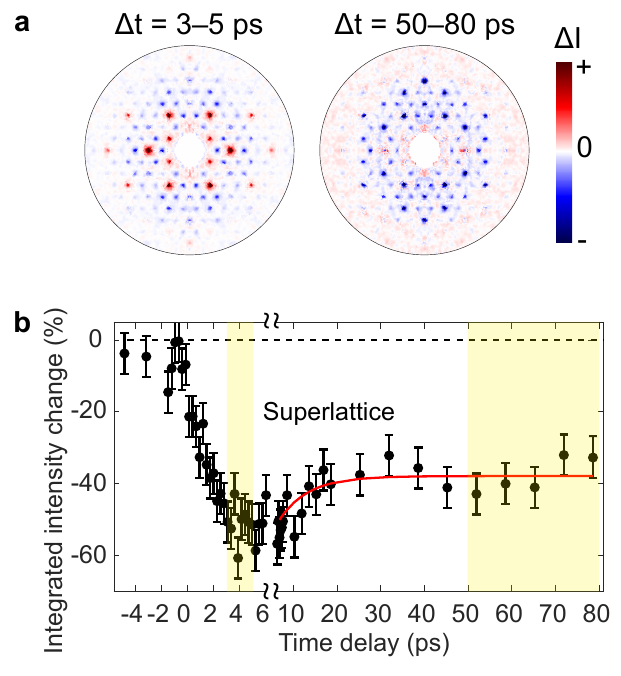}
    \caption{\textbf{Re-formation of superstructure and Debye-Waller dynamics. a} Representative symmetrised difference images for early (centered around $\approx\!~4$~ps) and late (around $\approx\!65$~ps) time delays. For improved signal-to-noise these are calculated by averaging difference patterns over time ranges, i.e. 3--5~ps for early and 50--80~ps for late time delays. For the late time delays, all Bragg peaks exhibit a loss of intensity indicative of a  Debye-Waller effect. \textbf{b} Dynamics  of the PLD satellite peaks over an extended time scale. Yellow shading indicates time ranges that were averaged to generate the difference images. A partial recovery is observed, fit with an exponential relaxation with offset (red line) with time constant ${\tau = 6.6 \pm 2.6}$ ps. The long-lived suppression indicates that the lattice system has thermalised at an elevated temperature.}
    \label{CDWrecovery}
\end{figure}

\vspace{3 mm}

\noindent {\bf Re-formation of lattice superstructure.} 
Following the melting of the trimer clusters, further structural dynamics ensues that is marked by the recovery of the ($3\times3$) superstructure due to lattice thermalisation. Figure~\ref{CDWrecovery}a shows transient difference maps, where the data has been averaged for two representative time ranges corresponding to early ($\Delta t=3\text{--}5$~ps) and late (50--80 ps) time delays. The pattern at early times exhibits the clear signature of the (${3\times3}$) trimer superstructure melting discussed above, while at later times such signature disappears and the pattern recovers the LT superlattice. The broad overall reduction of the peak intensities in this time range is that expected of a heated state with incoherent thermal atomic motions via the Debye-Waller effect \cite{Erasmus2012,Storeck_2020}. Between these time ranges, we note that diffraction peaks on one side of the pattern increase while those on the other side decrease, indicating mechanical buckling of the sample as it accommodates the structural phase change and lattice heating\cite{Zewail2008buckling,Lindenberg2015MoS2Buckling}. Difference images from these time ranges are shown in Supplementary Note 4.

As seen in Fig.~\ref{CDWrecovery}b -- which plots the time-evolution of the PLD side peaks over the entire measurement window -- the suppressed trimer order recovers $\approx\!20$\% of the original intensity with a $\approx\!7$ ps time constant. We note that this timescale may be influenced by simultaneous contribution from mechanical buckling. The overall suppression of the diffraction peaks relative to the LT ground state indicates the  thermalisation of the lattice degrees-of-freedom into a ``hot" (${3\times3}$) trimer superstructure at longer delays. We can estimate the maximum lattice thermalised temperature due to full thermalisation of the absorbed laser energy.
For the given 2.3~mJ cm$^{-2}$ fluence this results in a temperature of 177 K, based on the  1T’-\ce{TaTe2} heat capacity \cite{Sorgel2006} and the optical constants we measured of our 60 nm thick flake (see Supplementary Notes 8 and 9). We also performed experiments on the same sample at 1.75 mJ cm$^{-2}$ and 1.5 mJ cm$^{-2}$ fluence. The signature of structural phase transition was reproduced for these fluences with reduced magnitude (see Supplementary Note 9).    

\begin{figure*}[t]
    \includegraphics[width=15cm]{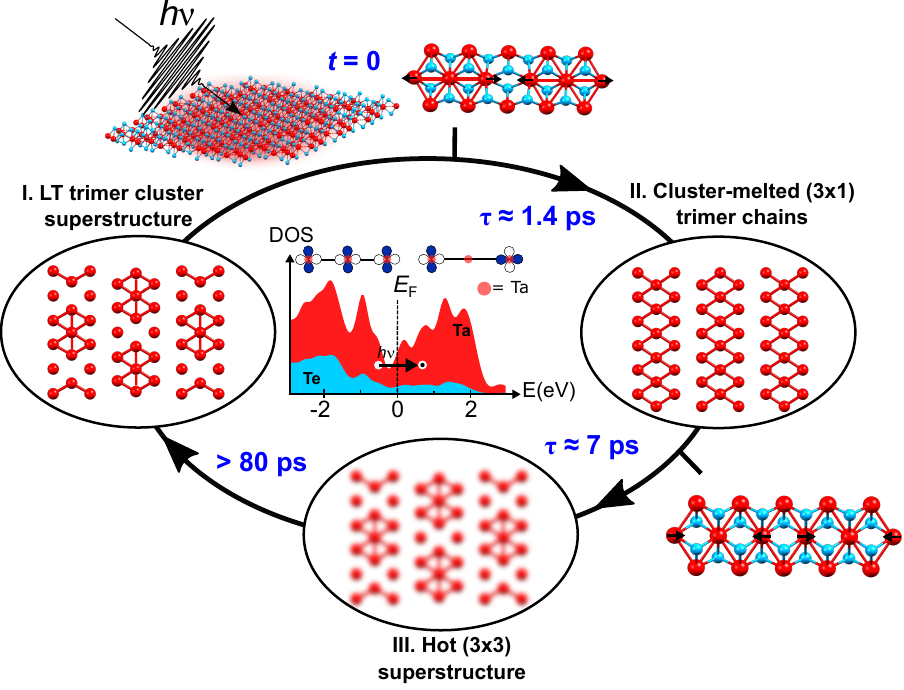}
    \caption{\textbf{Proposed stages of the  photo-induced dynamics in 1T'-\ce{TaTe2}}. The system starts in the LT phase with a (${3\times3}$) superstructure of Ta trimer clusters  (I). Near-IR excitation promotes carriers from bonding to non-bonding states via a charge-transfer excitation, which results in melting of the trimer clusters and leaves behind stripe-like trimer chains (II). Subsequently, the lattice subsystem thermalises, causing the (${3\times3}$) superstructure to re-establish and stablise at elevated temperatures (III). Inset: DFT-calculated density of states projected onto Ta and Te atoms, along with schematic optically-induced promotion from bonding to non-bonding Ta orbitals, which lie in the plane. The participating Ta trimer bonding and non-bonding (with a node in the middle) orbitals are shown.} 
    
    \label{simCompare}
\end{figure*}

\section*{Discussion}

\noindent Figure~\ref{simCompare} illustrates the proposed sequence of phases in the ultrafast dynamics of \ce{TaTe2}. Photo-excitation with intense pulses leads to the excitation of energetic carriers and picosecond melting of the low-temperature (${3\times3}$) trimer superstructure phase. The transient phase that results corresponds to a (${3\times1}$) trimer chain order, as inferred from the close overlap of the photo-induced change in the diffraction pattern with the fingerprint of the LT-HT transition. In the subsequent picoseconds, the lattice degrees-of-freedom thermalise, enhancing Ta-Ta bonds and thereby switching into a hot (${3\times3}$) superstructure state. The latter persists for extended times ($\gg80$~ps) until thermal diffusion transfers heat into the substrate. 

Density functional theory (DFT) calculations clarified the electronic states involved in the photo-excitation. We computed the projected density of states (DOS) and crystal orbital Hamilton population (COHP) of the relaxed LT and HT structures of \ce{TaTe2}, revealing the nature of occupied and unoccupied states, as shown in Fig.~\ref{simCompare} (see Supplementary Note 1 for details). In the LT phase, the lower-energy region of the valence band (i.e. below -4 eV) consists of Te $p$-states with a small contribution from Ta $d$-states, whereas the upper region of the band (i.e. above -2 eV) has mostly Ta character. Negative COHP values in the upper valence band indicate the bonding nature of Ta trimer states in this energy region. Trimer formation is enabled by partial charge transfer from Te to Ta involving $d_{\text{xz}}$ and $d_{\text{xy}}$ states which leaves uneven charges on Ta sites and enhanced Ta-Ta bonding \cite{Che18}. Meanwhile, conduction band states near the Fermi level belong to non-bonding states of Ta trimers, with anti-bonding states lying higher in energy ($\approx\!4$~eV above $\text{E}_F$).

Optical absorption in 1T’-\ce{TaTe2} involves mainly two kinds of dipole-allowed charge-transfer transitions (see Suppl. Note 1): promoting either Te $p$ to Ta anti-bonding states (type-I), or depopulating bonding states while populating non-bonding states of the $b$-axis trimers above the Fermi level (type-II). Photo-excitation around 1.2 eV chiefly involves the latter, which weakens the original charge-disproportionation between Ta sites within these trimers, thus triggering the ``melting" of the (${3\times3}$) order. The calculations also identified several strongly coupled phonons, including a mode around 2.7 THz involving Te motions and displacements of the Ta ions along the trimer axes that may be set into motion after optical charge transfer excitation. This predicts a cooperative mechanism for photo-induced trimer cluster dissolution in \ce{TaTe2}, which can be addressed in future diffraction studies with higher temporal resolution and time-resolved diffuse scattering to directly track the phonon modes participating in the transformation. 

Our study hence represents the first ultrafast measurement of \ce{TaTe2}, utilising short MeV electron bunches to resolve a picosecond atomic-scale melting of its intriguing trimer clusters and the subsequent thermalisation into a hot (${3\times3}$) superstructure phase. In these and other \ce{MTe2} systems (where M = transition metal), changes in the lattice structure are linked to anomalous changes in conductivity and magnetic susceptibility. The light-driven toggle and recovery seen here to occur between different lattice symmetries may thus enable applications, e.g. for ultrafast switching. Moreover, the associated trimer dynamics in this material opens the possibility for control of the related electronic modulations in \ce{TaTe2} on even faster time scales.

\section*{Methods}

\noindent {\bf Ultrafast Electron Diffraction.}
The UED experiments were performed  at the High Repetition-rate Electron Scattering (HiRES) beamline at Lawrence Berkeley National Laboratory (LBNL). The instrument exploits a one-of-a-kind technology developed at LBNL to provide unique beam properties for ultrafast structural dynamics studies, coupling relativistic electrons and high repetition rates. The results reported in the paper validate the technological breakthrough, which has the potential of broadening the scientific reach of ultrafast tools. Near-IR laser  pulses centred around 1030 nm wavelength and with $\approx\!350$~fs duration (full-width at half-maximum, FWHM) were used to photo-excite the sample. Synchronised electron pulses with 0.75 MeV kinetic energy (de Broglie wavelength $\lambda = 0.01$ \AA) were delivered to the sample for electron diffraction. Experiments were performed at 0.5~kHz repetition rate, recording 10 frames with the pump beam OFF and 10 frames with the  pump ON for each time delay. Each frame was recorded for an exposure time of 8 seconds. A home-built sample stage interfaced to a closed-cycle cryostat was used to cryogenically cool the samples down to 10~K. Laser and electron beams impinged on the \ce{TaTe2} sample surface, transmitting through a supporting silicon nitride window. The electron beam diameter at the sample was $\approx\!450~\mu$m FWHM, while the pump beam was adjusted to $\approx\!750~\mu$m FWHM to ensure homogeneous excitation across the probe volume. The beam charge was $\approx\!2.5$~fC corresponding to $1.6\times10^4$ electrons/pulse.

\vspace{2 mm}
\noindent {\bf \ce{TaTe2} samples.}
Single crystals of 1T'-\ce{TaTe2} were grown by the chemical vapour transport technique \cite{Ubaldini2013} at $700-800^{\text{\degree}}$C for 14 days using iodine as a transport agent. The crystals were repeatedly mechanically exfoliated with Scotch tape until optically transparent and then dry transferred onto 20-nm thick \ce{Si3N4} windows using polydimethylsiloxane (PDMS) stamps \cite{Castellanos_Gomez_2014}. The thickness of the \ce{TaTe2} flake was determined to be $\approx\!60$~nm using atomic force microscopy (AFM) (see Supplementary Note 2).

\vspace{2 mm}
\noindent {\bf Simulations, image analysis, and standard error.}
Details of DFT calculations, electron diffraction simulations, and diffraction image analysis methods are provided in Supplementary Notes 1, 3, and 4, respectively. The error bars in Figs. 2c and 3b represent the standard error for the given number of averaged frames. The corresponding standard deviation per frame is derived (for each peak set $\text{Bragg}\uparrow$, $\text{Bragg}\downarrow$, and superlattice) from the distribution of laser-off frames, multiplied by $\sqrt{2}$ to account for the total standard deviation when subtracting (laser on) - (laser off) frames.

\section*{Acknowledgements}
\noindent  We gratefully acknowledge Nord Andresen for developing the cryogenic sample stage and other excellent engineering efforts at HiRES, Paul Ashby for performing the AFM measurement, and Germ\'{a}n Sciaini for stimulating discussions. K.M.S., D.F., and R.A.K. acknowledge support for ultrafast materials UED studies by the Laboratory Directed Research and Development (LDRD) Program of Lawrence Berkeley National Lab under U.S. Department of Energy (DOE) Contract DE-AC02-05CH11231. Development and operation of the HiRES instrument (D.F., F.C) was supported by DOE under the same Contract No. Funding for D.B.D. was provided by STROBE: A National Science Foundation Science and Technology Center under Grant No. DMR 1548924. 
Work at the Molecular Foundry was supported by the DOE Office of Basic Energy Sciences under Contract No. DE-AC02-05CH11231. C.O. acknowledges support from the DOE Early Career Research Award program. A.R. gratefully acknowledges support through the Early Career LDRD Program of Lawrence Berkeley National Laboratory under DOE Contract No. DE-AC02-05CH11231. S.R. was supported through the Center for Non-Perturbative Studies of Functional Materials funded by the Computational Materials Sciences Program of DOE Office of Basic Energy Sciences, Materials Sciences and Engineering Division. This research used resources of the National Energy Research Scientific Computing Center, a DOE Office of Science User Facility supported by the DOE Office of Science under Contract No. DE-AC02-05CH11231. The financial support for sample preparation was provided by the National Science Foundation through the Penn State 2D Crystal Consortium-Materials Innovation Platform (2DCC-MIP) under NSF cooperative agreement DMR-1539916.

\noindent
\section*{Additional Information}
\noindent The authors declare no competing financial interests. \\
Correspondence and requests for materials should be addressed to R.A.K. (e-mail: kaindl@asu.edu) and D.F. (email: dfilippetto@lbl.gov). 
\noindent This work was published in Communications Physics 4, 152 (2021) available online at 
\url{https://doi.org/10.1038/s42005-021-00650-z}, and distributed under a CC BY license.  It is made 
available here as a reprint in author formatting under the same terms of the Creative Commons CC 
BY license. Copyright (c) 2021 The Authors, some rights reserved.

%\bibliographystyle{naturemag}
%\bibliography{main}% Produces the bibliography via BibTeX.

\providecommand{\noopsort}[1]{}\providecommand{\singleletter}[1]{#1}%

% switch to single column format
\onecolumngrid

% renew some of the commands
\renewcommand\bibname{Supplementary References} % change name of bibliography in the text 
\numberwithin{equation}{subsection} % add subsection number to equation
\renewcommand{\partname}{}
\renewcommand{\thesection}{\arabic{section}}
%\titleformat{\section}{\Large\bfseries}{Supplementary Note \thesection:}{0.75em}{}
\renewcommand{\thetable}{\arabic{table}}   % add S to table number
\renewcommand{\tablename}{\textbf{Supplementary Table}}
\renewcommand{\theequation}{S\arabic{section}.\arabic{equation}}  % add S to equation number
\renewcommand{\figurename}{\textbf{Supplementary Figure}}
\graphicspath{{figures/}} % path to figures directory
\setcounter{figure}{0} % reset the figure counter
\setcounter{page}{0}

\titleformat{\section}{\large\bfseries}{Supplementary Note \thesection:}{0.75em}{}

\newpage

\begin{center}
\end{center}
\begin{center}
\large{Supplementary Information}\\
\end{center}
\begin{center}
\Large{\textbf{Ultrafast optical melting of trimer superstructure in layered 1T'-\ce{TaTe2}}}\\ 
\end{center}

\begin{center}
{Khalid M. Siddiqui, Daniel B. Durham, Frederick Cropp, Colin Ophus, Sangeeta Rajpurohit, Yanglin Zhu, Johan D. Carlstr\"{o}m, Camille Stavrakas, Zhiqiang Mao, Archana Raja, Pietro Musumeci, Liang Z. Tan, Andrew M. Minor, Daniele Filippetto, and Robert A. Kaindl}
\end{center}

\newpage

\vspace{5mm}

\section{Density functional theory}
\label{sec:DFT}

\begin{figure}[h!]
    \centering
    \includegraphics[width=13cm]{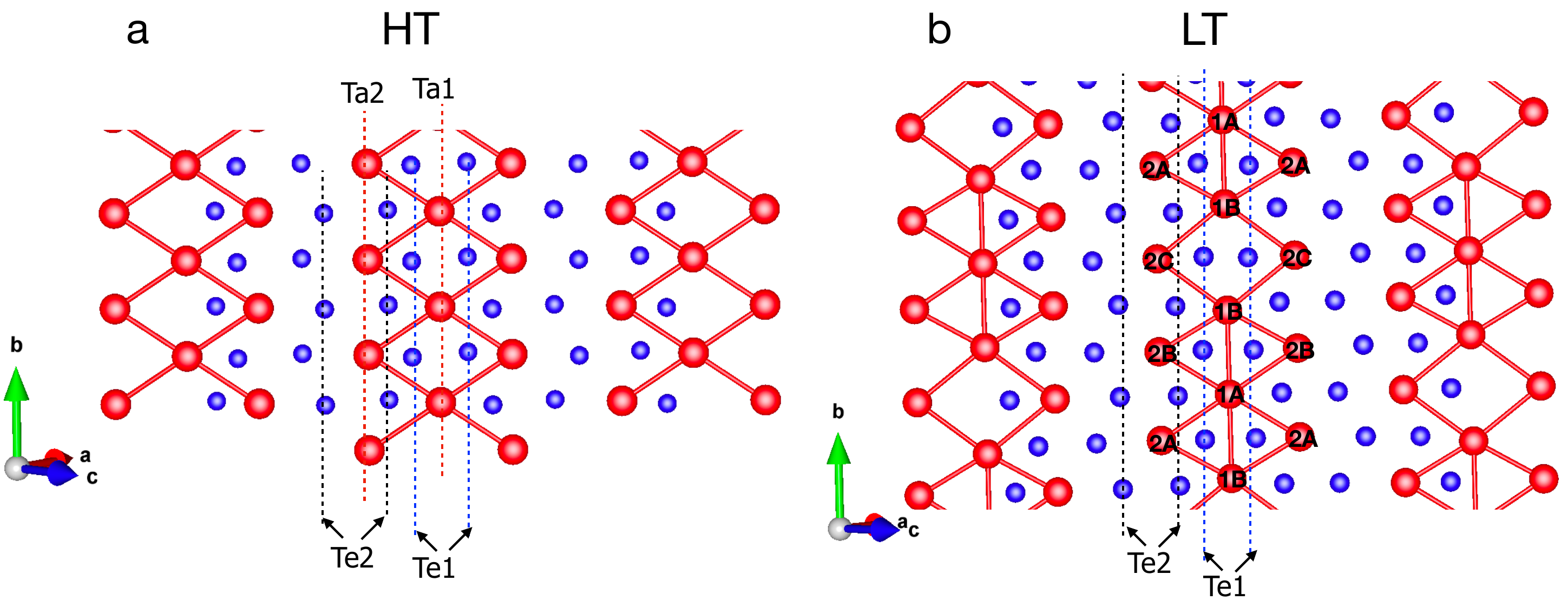}
    \caption{\textbf{a, b} DFT relaxed atomic structures of HT and LT phase viewed within a single layer. Two inequivalent Ta ions in HT are indexed as Ta1 and Ta2. The Te ions within $xy$ plane around Ta1 ions are indexed as Te1, the respective Te ions around Ta2 are indexed as Te2. The four inequivalent Ta ions in LT are indexed as 1A, 1B, 2A, and 2B. The Te ions in the $xy$ plane around 1A and 1B Ta ions are labelled as Te1. For 2A and 2B Ta ions, the same Te ions are labelled as Te2.}
    \label{fig:RelaxedStructures}
\end{figure}

To understand the melting of lattice distortions in \ce{TaTe2}, we investigated the nature of electronic states involved in the charge-transfer transitions during photo-excitation. We performed DFT calculations with the \textit{Vienna ab initio Simulation Package} (VASP) using projected-augmented wave basis set \cite{Blochl1994,Kresse1996,Kresse1999} and experimental lattice parameters \cite{Sorgel2006}. We choose a plane wave cut-off energy of 350 eV. A \textit{k}-grid of size ($8\times8\times8$) and  ($4\times4\times4$) is used for the LT and HT ionic relaxation, respectively. For the ionic relaxation, we use the Perdew-Burke-Ernzerhof generalised gradient approximation (GGA) with a force convergence criterion of 1meV/$\textnormal{\AA}$.

The HT and LT relaxed structures, shown in Supplementary Figure~\ref{fig:RelaxedStructures}, are consistent with the 
experimental atomic structure with ($3\times1$) and ($3\time3$) superstructures, respectively. The HT structure has two inequivalent Ta ions which forms the centre (Ta1) and edges (Ta2) of the trimers in the double zig-zag chain. The LT phase has four inequivalent Ta ions (1A, 1B, 2A and 2B). Every Ta ion is surrounded by an octahedron of Te ions.

\begin{figure}[h!]
    \centering
    \includegraphics[width=13cm]{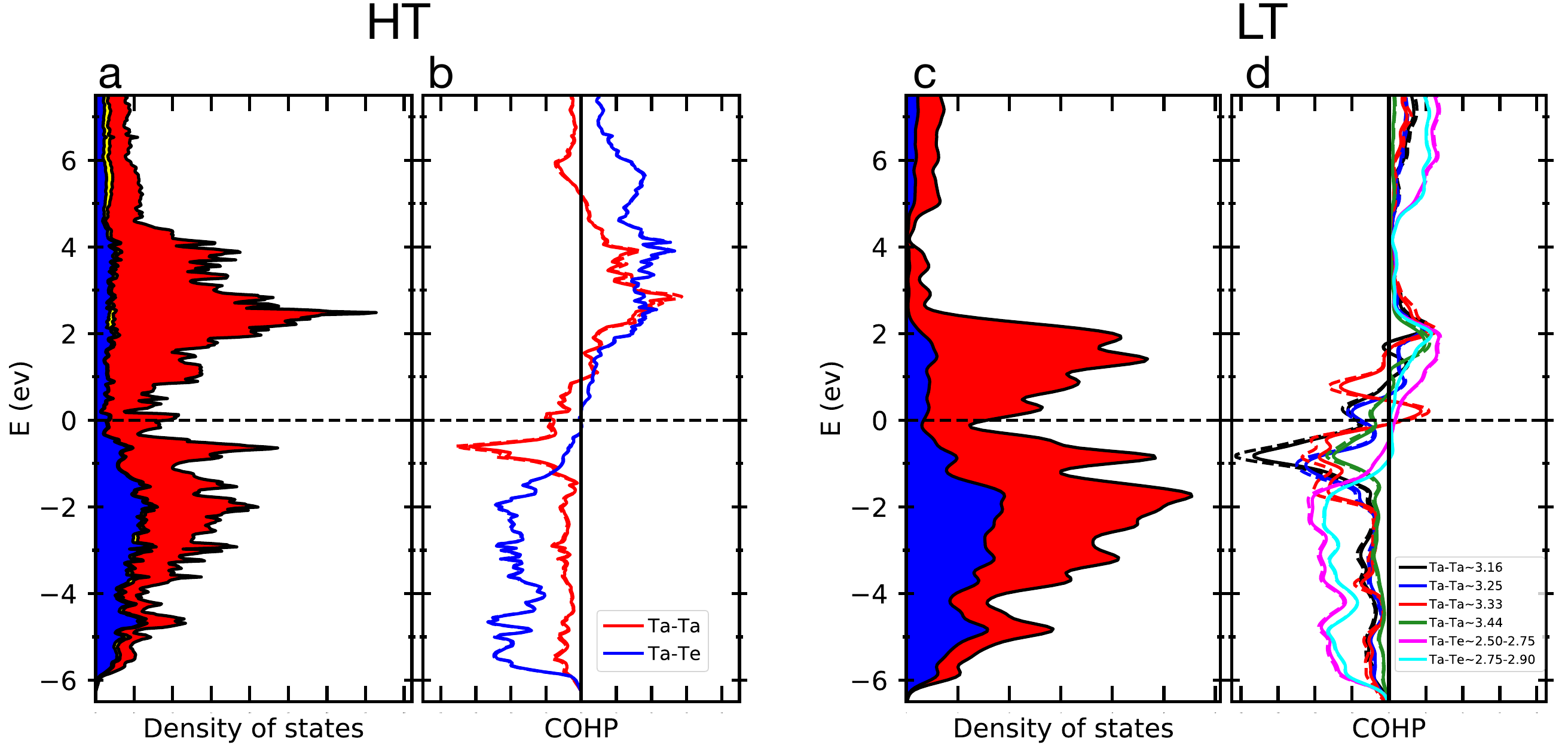}
    \caption{\textbf{a, c} Density of states of \ce{TaTe2} structure projected on Ta d-states (red) and Te s-states (yellow) and p-states (blue) for HT and LT relaxed structures. The Fermi-level is shown by horizontal dashed line. \textbf{b, d} COHP for HT and LT relaxed structures. For LT, corresponding bond lengths of the selected bonds are listed in the legend (in Angstroms).}
    \label{fig:DOS}
\end{figure}

To investigate the energy-resolved bonding nature between ions, we performed a crystal orbital Hamilton population (COHP) analysiswith the local orbital basis suite towards electronic structure reconstruction (LOBSTER) code. The Ta-Te COHP for the HT phase in Figure~\ref{fig:DOS} is the averaged over all Ta-Te pairs. For the Ta-Te COHP in the LT phase, we categorise Ta-Te pairs into two groups with inter-atomic distance in the range 2.50-2.75 $\textnormal{\AA}$ and 2.75-2.95 $\textnormal{\AA}$. 

The projected density of states and COHP of the relaxed HT and LT structures are shown in Supplementary Figure~\ref{fig:DOS}. Both structures are metallic with overlapping bands at the Fermi level. However compared to HT, the LT structure has larger density of states at the Fermi level. The lower energy region of the valence band of both structure consists of Te \textit{p}-states with a small weightage from Ta \textit{d}-states. In addition, the negative COHP between Ta1(Ta2) and Te1(Te2) states confirms the bonding nature of these states. The corresponding anti-bonding states, with positive COHP, with largely Ta contribution form the conduction band. 

\begin{table}[!hbt]
\caption{Integrated COHP (ICOHP) between Ta atoms at the Fermi level calculated for LT and HT phase. ICOHP values in bracket are for opposite spin direction.}
\begin{center}
    \begin{tabular}{|c|c|c|c|}
    \hline\hline
     Phase & Bond type & Distance ($\textnormal{\AA}$) & ICOHP   \\
     \hline
    \multirow{7}{*}{LT} & 
    Ta(1A)-Ta(2A) & 3.256 & -0.957(-0.952)  \\
   & Ta(1B)-Ta(2A) & 3.160 & -1.281(1.282)  \\  
   & Ta(1B)-Ta(2B) & 3.445 & -0.622(-0.625) \\
   & Ta(1B)-Ta(1A) & 3.331 & -0.989(-1.012)  \\
   & Ta(2A)-Ta(2A) & 3.517 & -0.693(-0.686) \\
   & Ta(2B)-Ta(2B) & 3.684 & -0.363(-0.366) \\ 
   & Ta(1B)-Ta(1B) & 4.220 & -0.020(-0.029) \\
    \hline\hline
    \multirow{2}{*}{HT} & 
     Ta1-Ta2 & 3.282 &  -0.491(-0.491)\\
   & Ta1-Ta1/Ta2-Ta2 & 3.637 & -0.209(-0.209)\\ \hline\hline
    \end{tabular}
\label{tab:cohp_tab}
\end{center}
\end{table}
 
The upper region of the valence band has mostly Ta contribution. The negative COHP between Ta1 and Ta2 reveals the bonding nature of the states. We attribute the upper region of the valence band to the bonding states of Ta trimers, as suggested by previous studies \cite{chen2018trimer}. The respective non-bonding and anti-bonding states of the trimers lie high in energy forming the conduction band. The integrated COHP (ICOHP) between Ta atoms at Fermi level is summarised in Supplementary Table~\ref{tab:cohp_tab}. The average Ta-Ta COHP along trimer axis in LT phase has greater magnitude compared to HT phase. This indicates the Ta-Ta bonding is stronger in LT phase. ICOHP suggests the Ta-Ta bonding along \textit{b} direction in HT phase is very small or negligible. However, the LT phase has a significantly stronger Ta-Ta bonding, confirmed by higher ICOHP, along $b$ direction. 

\begin{figure}[h!]
    \centering
    \includegraphics[width=13cm]{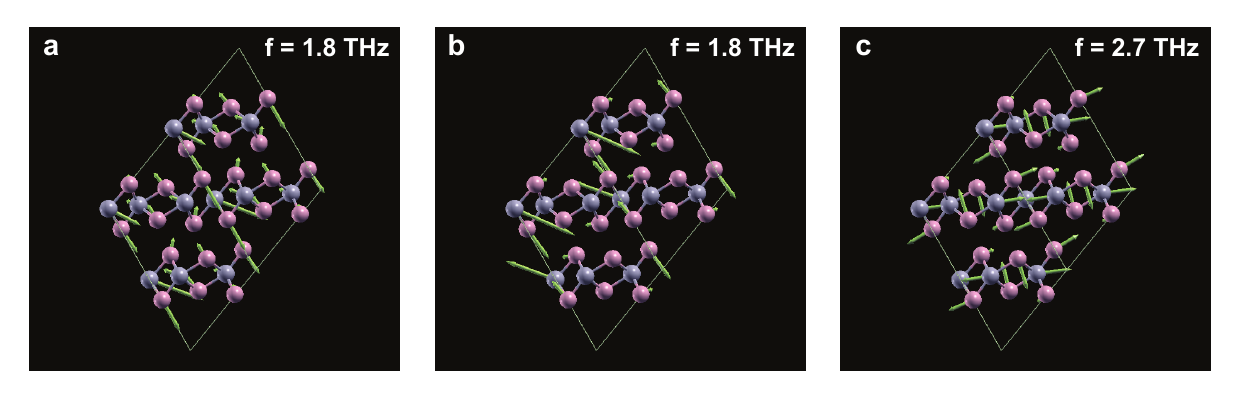}
    \caption{Three optical phonon modes found to strongly couple to non-bonding Ta electronic levels which dominate in the range from 1.2 eV above to 1.2 eV below the Fermi energy accessible by our pump photons, as obtained from excited-state DFT calculations. Such states are excited via the type-II transitions described in the text. \textbf{a} in-plane in-phase Ta mode with $B_u$ symmetry. \textbf{b} in-plane out-of-phase Ta mode with $B_u$ symmetry. \textbf{c} in-plane Ta and out-of-plane Te displacements mode with $A_g$ symmetry. Ta atoms are shown in blue, Te in purple. Green arrows: relative displacement vectors.
    }
    \label{fig:PhononModes}
\end{figure}

The optical absorption in 1T'-\ce{TaTe2} involves mainly two kinds of dipole-allowed charge-transfer transitions: between Te \textit{p}-states to Ta anti-bonding states (type-I); and from bonding states to the non-bonding states  of the Ta trimers (type-II). Photo-excitation around 1.2 eV predominantly couples to non-bonding Ta-trimer states (type-II transition) -- resulting in a charge transfer that weakens the original charge-disproportionation within the trimers and thus triggers a “melting" of the (${3\times3}$) order with associated lattice dynamics. Higher-energy photons in the UV instead will chiefly induce type-I transitions, which redistribute electrons from Te to Ta and do not affect the original PLD of \ce{TaTe2} that is attributed to unequal charges among the Ta ions. Thus, more energetic pump photons may not induce the transient phase transition as efficiently.

Vibrational modes and electron-phonon couplings were calculated within Density Functional Perturbation Theory~\cite{giustino_electron-phonon_2017}, as implemented within the Quantum Espresso software package~\cite{giannozzi_quantum_2009}. We solve for phonon modes at the $\Gamma$-point, and calculate electron-phonon coupling matrix elements for bands within 1.2 eV of the Fermi level. Supplementary Fig.~\ref{fig:PhononModes} shows three modes found to strongly couple to the electronic levels from 1.2 eV above to 1.2 below the Fermi level, as accessible by our pump photons. These vibrational modes at 1.8~THz ($B_u$ symmetry) and 2.7~THz ($A_g$ symmetry) can in principle be coherently triggered after the optical excitation. Due to symmetry, the $A_g$ mode is Raman active and of low enough symmetry to participate in displacive excitation of coherent phonons (DECP), while the $B_u$ modes should couple neither via impulsive Raman nor DECP. From this, we expect that coherent phonon generation may involve impulsive stimulated Raman scattering (and possibly DECP) of the $A_g$ mode around 2.7 THz. Additional insight, however, necessitates future experiments with sufficiently high time resolution to resolve the oscillations.

\newpage
\vspace{5mm}
\color{black}
\section{\ce{TaTe2} sample preparation and characterisation}
\label{sec:Sample}

A single crystal of \ce{TaTe2} was synthesised through the chemical vapour transport (CVT) method  [1,2]. The Ta, Te powder, and 20 mg \ce{I2} were mixed with a stoichiometric ratio and loaded into a quartz tube (10 cm inside diameter, 18 cm length), then sealed under vacuum. The quartz tube was heated up by a double-zone furnace, where the hot end and the cold end were set at $850{\text{\degree}}$C and $750{\text{\degree}}$C, respectively. The temperature gradient was kept for 14 days. Then the furnace was shut down, and the quartz tube was naturally cooled down to room temperature. The black plate-like crystals were found at the cold end of the quartz tube.

\begin{figure}[h!]
    \centering
    \includegraphics[width=16cm]{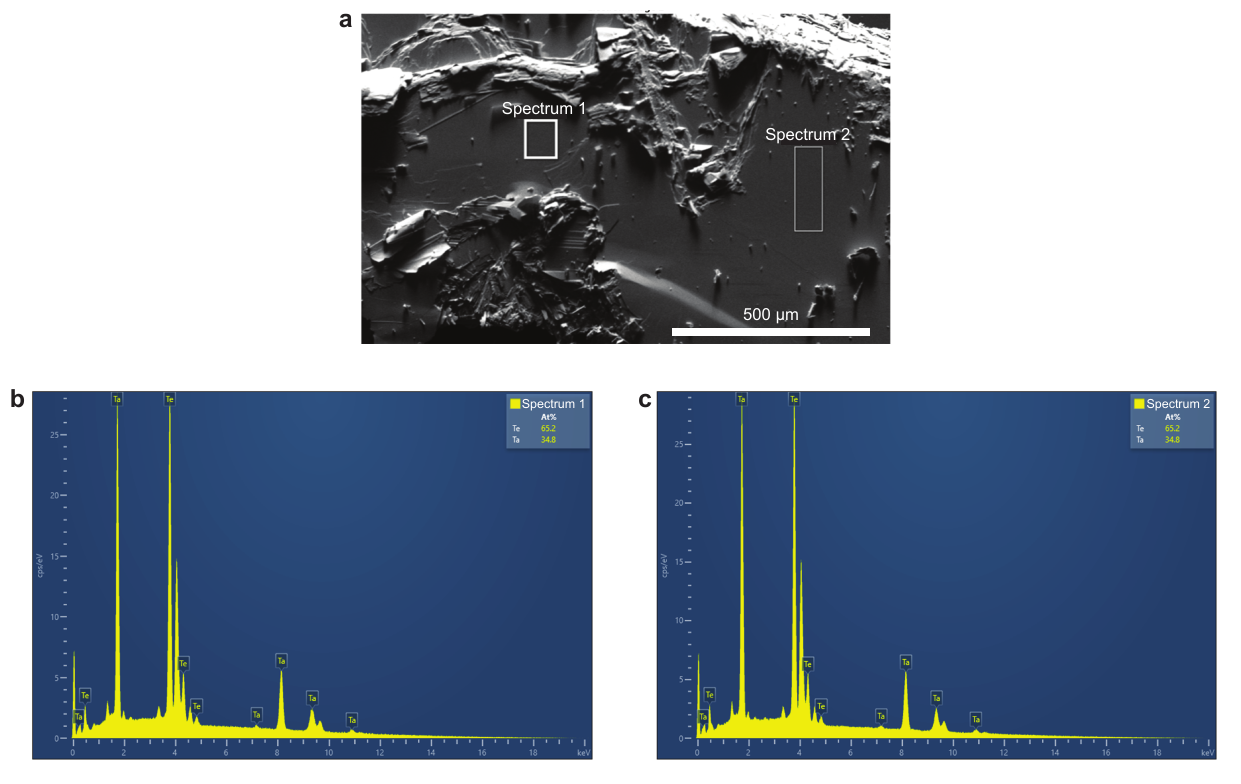}
    \caption{Composition characterisation of CVT-grown \ce{TaTe2} crystal using EDS in a scanning electron microscope. \textbf{a} SEM micrograph of the crystal, highlighting two regions from which EDS spectra are obtained. \textbf{b-c} Average EDS spectra obtained while rastering the electron beam over the regions indicated in the micrograph.}
    \label{fig:eds}
\end{figure}

The composition of the synthesised crystals was confirmed to be close to \ce{TaTe2} using energy-dispersive x-ray spectroscopy (EDS), shown in Supplementary Figure~\ref{fig:eds}. The composition in two regions separated by nearly a millimeter is found to be the same, suggesting uniform composition across the sample.

The resistivity of the material was measured as a function of temperature, using a crystal from same batch used for the UED experiments. The results are shown in Supplementary Figure \ref{fig:ResitivityPlot}. The plot covers both cooling and warming cycles. A drop in resistivity is observed around $T_c$ = 174 K which marks the onset of the phase transition to the trimer superstructure phase. The step-like drop and hysteresis are indicators of a first-order phase transition.

\begin{figure}[h!]
    \centering
    \includegraphics[width=9cm]{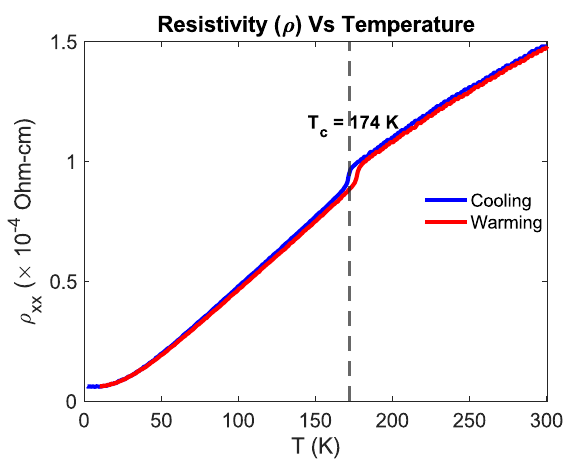}
    \caption{Temperature-dependent in-plane resistivity $\rho_{xx}$ as measured for a \ce{TaTe2} single crystal from our batch of samples grown by the chemical vapor transport method.}
    \label{fig:ResitivityPlot}
\end{figure}

The exfoliated flake for UED experiments was mounted on a \ce{Si3N4} membrane suspended across a 30 $\mu$m x 30 $\mu$m square window in a Si support chip produced by Norcada. A white light microscope image is shown in Supplementary Figure~\ref{fig:sample}. The reflectance contrast differs between the Si-supported region of the flake and the region only supported by the \ce{Si3N4} membrane, suggesting the sample is thin enough to permit some optical transmission. Optical properties in the red to NIR region (700-1100 nm) are characterised in more detail in Supplementary Note~\ref{sec:Reflectivity}.

\begin{figure}[h!]
    \centering
    \includegraphics[width=13cm]{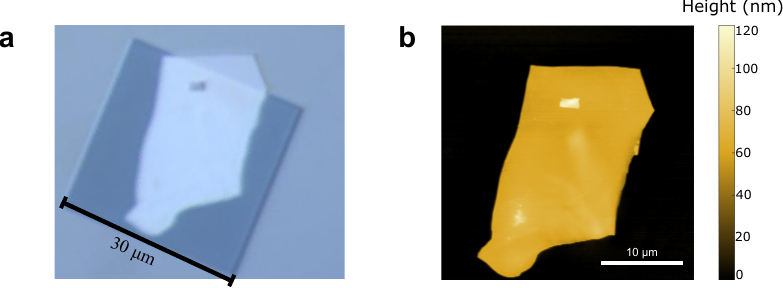}
    \caption{\textbf{a} White light microscope image of \ce{TaTe2} flake on 20 nm silicon nitride window using reflected light. \textbf{b} AFM topography map of \ce{TaTe2} flake.}
    \label{fig:sample}
\end{figure}

We used tapping-mode atomic force microscopy to map the topography of the sample (Supplementary Figure~\ref{fig:sample}). The thickness of the sample was determined to be 61.5 nm. Some rippling of the sample was observed on micrometer length scales, with a standard deviation of 5.8 nm. The influence of this rippling on the diffraction patterns is discussed in Supplementary Note~\ref{sec:DiffSims}. 

\begin{figure}[h!]
    \centering
    \includegraphics[width=9cm]{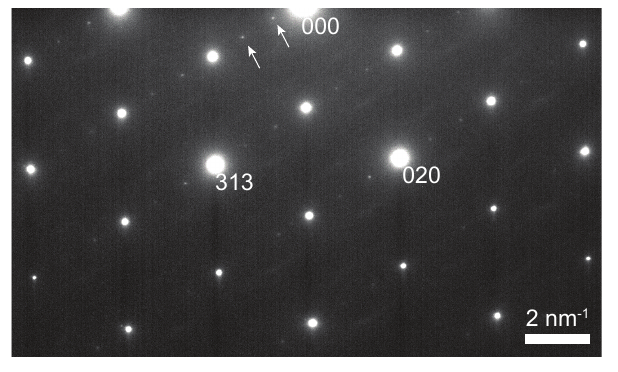}
    \caption{Selected area diffraction pattern from the \ce{TaTe2} flake measured in a 300 kV TEM at room temperature. Weak satellite peaks are observed between the primary diffraction peaks, including the two highlighted by the white arrows.}
    \label{fig:TEM}
\end{figure}

The crystallinity of the \ce{TaTe2} flake was characterised before UED measurements using selected area diffraction (SAED) in TEM using a 300 keV electron beam at room temperature, including the diffraction pattern shown in Figure~\ref{fig:TEM}. In addition to the intense peaks corresponding to the main lattice Bragg spots observed with the beam at HiRES (750 keV), we observe satellite peaks at room temperature with about 1000 times lower intensity than adjacent primary diffraction peaks. These peaks appear perpendicular to the \textit{b} axis, along the \textit{a*} direction indicated in Figure 1, with a \textbf{q} vector of 1/3 (111). Such peaks have been observed in prior TEM studies of \ce{TaTe2}\cite{wei2017TEMTaSeTe,wang2020TEMMX2}. They are not predicted to appear in kinematical simulations: instead, their presence has been attributed to limited coherence length of the PLD along the stacking direction\cite{wei2017TEMTaSeTe} and may involve contribution from random thermal atomic motion and multiple scattering. 

\newpage
%\color{black}
\section{Electron diffraction simulations of contributions to diffraction peak dynamics}
\label{sec:DiffSims}
\begin{figure}[h!]
    \centering
    \includegraphics[width=16cm]{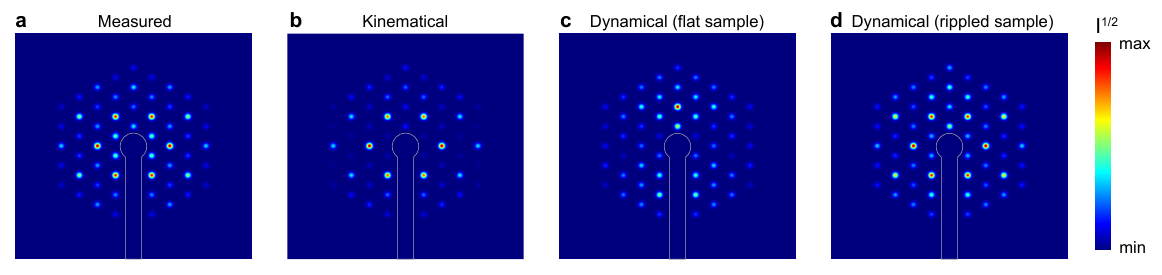}
    \caption{Comparison of kinematical and dynamical diffraction methods to the measured UED pattern at 300 K. \textbf{a} Symmetrised measured UED pattern. \textbf{b} Kinematical diffraction pattern for HT phase. \textbf{c,d} Dynamical diffraction pattern computed using multislice method considering a flat sample and a sample with gaussian distribution of tilt angles with $\sigma_{\theta}$ = 30 mrad. All simulated patterns show the square root of the intensity.}
    \label{fig:diffSimTypes}
\end{figure}

Kinematical diffraction simulations that consider the limit that each electron undergoes no more than one scattering event were carried out. In this case, the diffraction peak intensities for a single crystal are largely determined by the kinematical structure factors, with modifications due to factors such as sample thickness and incident electron energy. We use the SingleCrystal commercial software tool to compute kinematical electron diffraction patterns\cite{palmer2015CrystalMaker}. A computed pattern for the HT phase is shown in Supplementary Figure~\ref{fig:diffSimTypes}b.

Because of the high atomic numbers of the constituent Ta (Z = 73) and Te (Z = 52) atoms as well as the 60 nm thickness, electrons are likely to undergo multiple scattering events as they pass through this sample. To investigate the influence of multiple scattering, we used the multislice approach to compute dynamical electron diffraction patterns. In this approach, the electron beam interaction with the sample is modeled as sequential interaction with and propagation through slices of projected electrostatic potential of the atoms. The envelope of the electron wave function passing through the material can be obtained from the Schr\"{o}dinger equation for fast electrons as\cite{kirkland1998advanced}:
\begin{equation}
    \frac{\partial{\psi(\mathbf{r})}}{\partial{z}}=\bigg[\frac{i\lambda}{4\pi}\nabla_{xy}^{2}+i\sigma V(\mathbf{r})\bigg]\psi(\mathbf{r})
\end{equation}
Here, $\psi$ is the electron wave function, $\lambda$ is the electron de Broglie wavelength, $\sigma$ is the interaction parameter, and V is the electrostatic potential of the material. The two terms on the right-hand side correspond to propagation (left term) and interaction (right term). Within a finite slice, these two terms can be applied separately as operators\cite{ophus2017prism}. First, an interaction operator is applied in real space:
\begin{equation}
    \psi_{p+1}(\mathbf{r})=\psi_{p}(\mathbf{r})e^{i\sigma V_{p}^{2D}(\mathbf{r})}
\end{equation}
where $V_{p}^{2D}(\mathbf{r})$ is the projected electrostatic potential within the slice. We use the parameterised projected potentials for Ta and Te atoms as determined by Kirkland \cite{kirkland1998advanced}. Second, a propagation operator is applied in reciprocal space:
\begin{equation}
    \psi_{p+1}(\mathbf{q})=\psi_{p}(\mathbf{q})e^{-i\pi\lambda|\mathbf{q}|^{2}t}
\end{equation}

We start with a plane wave beam (spatially uniform envelope) and advance the beam through each slice by applying the interaction and propagation operator until it reaches the end of the material. The diffraction is then obtained from the Fourier transform of the exiting envelope:
\begin{equation}
    I(\mathbf{q}) = |\mathcal{F}(\psi(\mathbf{r}))|^{2} = |\psi(\mathbf{q})|^{2}
\end{equation}
For the simulations shown here, we use a simulation box that is 1.92 nm long along the b axis and 1.09 nm along the a* axis (axes defined in Figure 1), with the thickness of the crystal divided into 0.223 nm slices. 

A computed pattern using the multislice method for a 60 nm thick film of the HT phase is shown in Supplementary Figure~\ref{fig:diffSimTypes}c. Strong dynamical effects are observed, leading to asymmetry in the pattern and dramatically different diffraction peak intensities than obtained via kinematical simulation. 

However, this calculation assumes a plane wave incident beam incident on a perfectly flat sample. As the AFM measurement in Supplementary Note~\ref{sec:Sample} demonstrates, rippling is present across the film, leading to an angular distribution of the order of several to tens of mrad. Because the electron probe is 100s of micrometers in size, the entire range of tilt angles is sampled simultaneously. We find the peak intensity distribution is sensitive sub-mrad changes in the sample tilt in these simulations due to strong interference effects between the scattered beams. To achieve more realistic behaviour, we must average the diffraction over the distribution of tilt angles present in the sample.

To account for these effects, we perform dynamical simulations for many tilt angles and perform a Gaussian-weighted average of the intensities from these tilts. For a 60 nm thick film, we find that a distribution of tilt angles with $\sigma_{\theta} = 30$ mrad provides the best agreement with the measured diffraction patterns (shown in Supplementary Figure~\ref{fig:diffSimTypes}d). This is larger than the $\sigma_{\theta} = 6.5$ mrad we calculate from the AFM topography map: however, this value is an underestimate as nanoscale topographic features are smoothed out in this large area scan (pixel size = 118 nm). By averaging over many tilt angles, the distribution of peak intensities is smoothed out and symmetrised across the pattern. Patterns simulated in this way show better agreement with the measured patterns than either single-tilt plane-wave simulations or kinematical diffraction simulations. Quantitative agreement could be further improved, for instance, by knowledge of the atomic displacement parameters and the exact distribution of tilt angles.

\begin{figure}[h!]
    \centering
    \includegraphics[width=13cm]{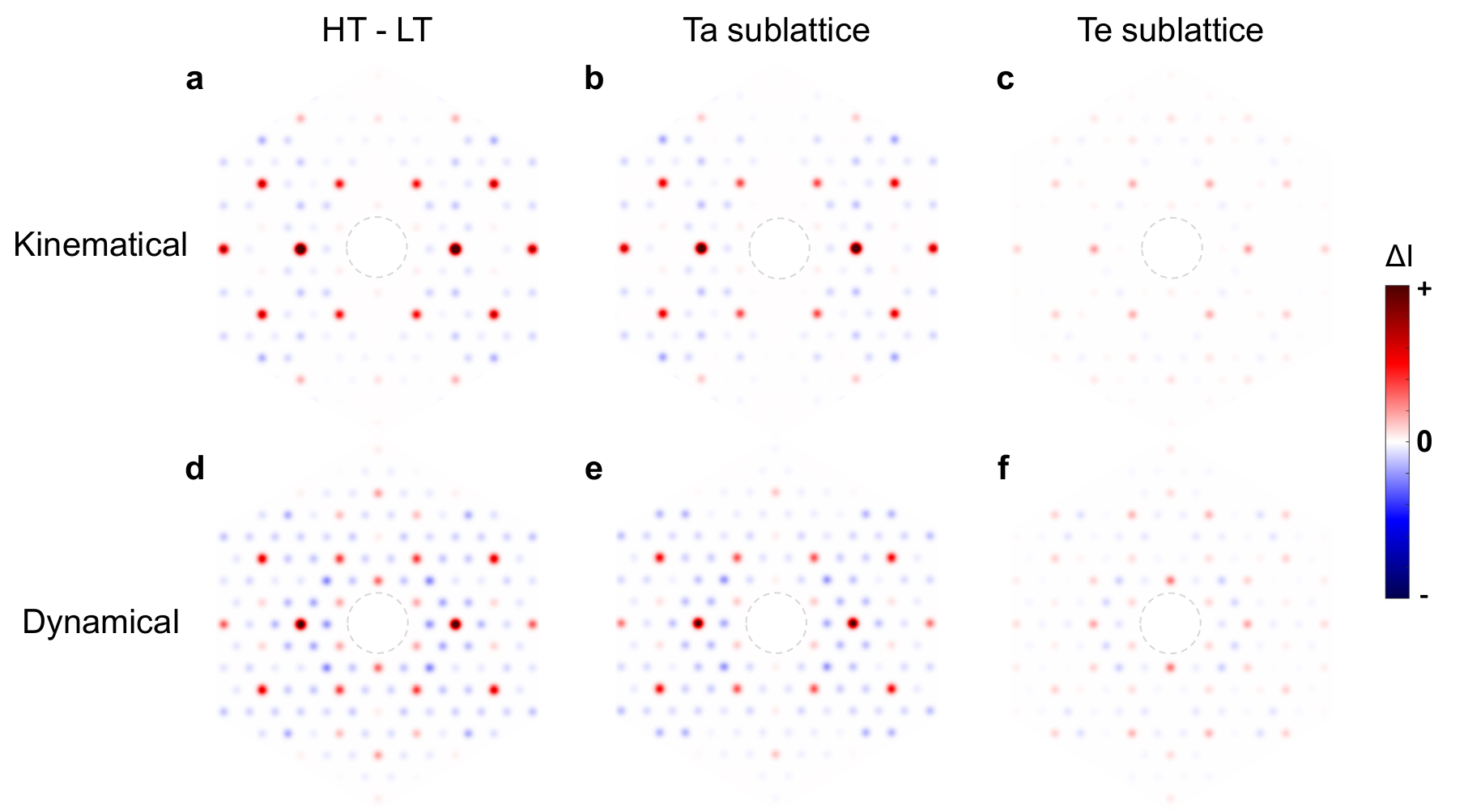}
    \caption{\textbf{Simulated difference patterns for the LT to HT transition.} \textbf{a-c} Kinematical simulations for both Ta and Te atom motions, Ta motions only, and Te motions only. \textbf{d-f} Multislice simulations with $\sigma_{\theta}$ = 30 mrad for both Ta and Te atom motions, Ta motions only, and Te motions only. The dashed circles at the centre of each difference image indicate the central part of the beamstop for ease of comparison with measured difference patterns.}
    \label{fig:diffSimSublattices}
\end{figure}

We employ both kinematical and dynamical diffraction simulations to investigate the contribution of atomic motions to changes in diffraction patterns of \ce{TaTe2}. The structural phase transition in \ce{TaTe2} involves multiple 3D atomic motions, including weakening of Ta-Ta bonds and reorganization of the Te chalcogens. In principle, one could separate the contributions of these motions and determine their timescales to understand whether they occur cooperatively or sequentially. For example, in $\mathrm{TaSe_{2-x}Te_x}$ it was found that Ta and chalcogen in-plane motions could be readily separated; Ta motions caused all primary diffraction peaks to rise, while Te motions caused some to rise and others to fall\cite{li2019ultrafast}. 

Difference images between simulated diffraction patterns for the LT and HT phase, as well as hypothetical structures where only the Ta or Te atoms move from LT to HT positions, are shown in Supplementary Figure~\ref{fig:diffSimSublattices}. We observe that the diffraction peak changes upon transformation from LT to HT phase (Supplementary Figure~\ref{fig:diffSimSublattices}a,d) are mostly accounted for by the Ta atom motions (Supplementary Figure~\ref{fig:diffSimSublattices}b,e), and that this motion alone causes a mixture of positive and negative changes in the primary diffraction peaks much like that observed in our UED experiment. On the other hand, changes due to Te atoms (Supplementary Figure~\ref{fig:diffSimSublattices}c,f) are smaller and somewhat different from those found for the Ta atoms. These simulations show that the overall symmetry change of the trimer superstructure within the highly distorted, monoclinic crystal structure causes the mixture of positive and negative diffraction peak changes observed in our experiment. Dynamical scattering modifies the magnitude and, for a few reflections, the sign of the changes, but is not the root cause of the mixed signs.

\begin{figure}[h!]
    \centering
    \includegraphics[width=13cm]{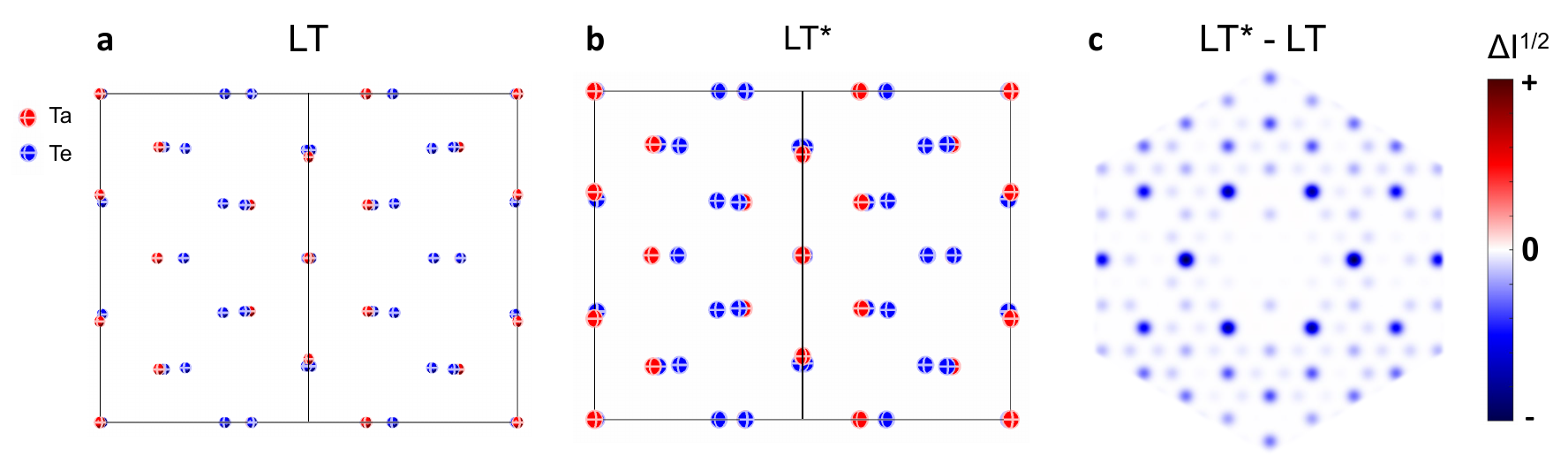}
    \caption{\textbf{Illustration of the effect of lattice heating on the electron diffraction pattern.} \textbf{a,b} Model of the low-temperature \ce{TaTe2} unit cell before thermalization (LT) and after thermalization (LT*). Thermal ellipsoids correspond to magnitude of incoherent atomic motions. \textbf{c} Difference image between LT and LT* diffraction patterns calculated using kinematical diffraction simulations. The square root of the difference is shown to more clearly observe the changes in the superlattice peaks.}
    \label{fig:diffKinHeating}
\end{figure}

To qualitatively illustrate thermal reduction of diffraction peak intensities, to which we attribute behavior in our UED patterns at longer time delays, we performed kinematical diffraction simulations including a "thermal ellipsoid." This ellipsoid represents the magnitude of random atomic displacements in the material. The difference image between diffraction patterns simulated with small and large thermal ellipsoids is shown in Supplementary Figure~\ref{fig:diffKinHeating}. The small thermal ellipsoids are those obtained from x-ray diffraction measurements of the LT phase\cite{Sorgel2006} whereas the larger ellipsoids are chosen to be twice as large for simplicity. Indeed, all diffraction peaks decrease upon heating, providing a qualitative explanation for the largely negative peak changes observed from 50 to 80 ps in the UED experiment. We note that for accurate quantitative modeling of the influence of thermalisation on the diffraction pattern, a good model of the atomic displacement magnitudes and accounting for dynamical scattering is needed.

\newpage

\section{Difference images for selected time ranges}
\label{sec:DiffImages}

\begin{figure}[h!]
    \centering
    \includegraphics[width=16cm]{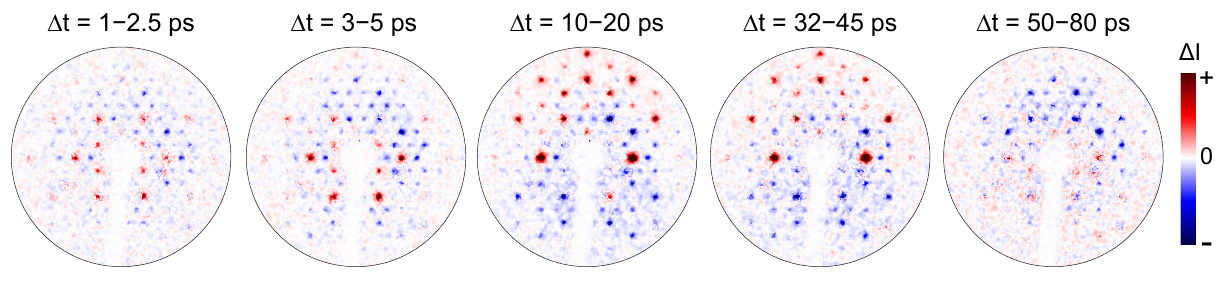}
    \caption{Photo-induced changes in the raw, unsymmetrised diffraction pattern at selected time ranges. For improved signal-to-noise these are calculated by averaging difference patterns over the time ranges listed above the patterns.
  }
    \label{fig:TimeDependentDiff}
\end{figure}

The photo-induced changes in the diffraction pattern at early (t$\approx\!~1.75$~ps and 4~ps), intermediate (t $\approx\!~10$~ps and 38.5~ps), and late (t $\approx\!~65$~ps) time delays are shown in Supplementary Figure~\ref{fig:TimeDependentDiff}. After several picoseconds, significant asymmetry in the changes is observed, with diffraction signals mainly increasing in the upper half of the pattern and mainly decreasing in the lower half. Intrinsic changes in the material would typically affect peaks on opposite sides equally, as their intensities are related by Friedel's law. These asymmetric changes, on the other hand, we attribute to mechanical buckling of the sample, causing the sample to tilt relative to the electron beam. Such an induced tilt misalignment would reduce the deviation from the Bragg condition of reflections on one side of the pattern relative to the other, consistent with the observed behavior. 

This buckling could be caused by a combination of structural phase transition of the \ce{TaTe2} flake and lattice heating, both of which cause volumetric expansion that would strain the flake. The asymmetric changes subside at the 50--80 ps time range, suggesting local equilibrium is then re-established. 

This asymmetric behaviour occurs simultaneously with the initial stages of PLD recovery, complicating isolation of the recovery timescale. Such an effect could be reduced in future studies if thinner flakes can be achieved, for which the diffraction peak intensities are less sensitive to sample tilt and photo-induced strain would likely be reduced. Nonetheless, these observations motivate further development of dynamical diffraction analysis techniques to account for contributions such as tilt misalignment, multiple scattering, and sample heating in UED experiments of strongly scattering samples.

\begin{figure}[h!]
    \centering
    \includegraphics[width=9cm]{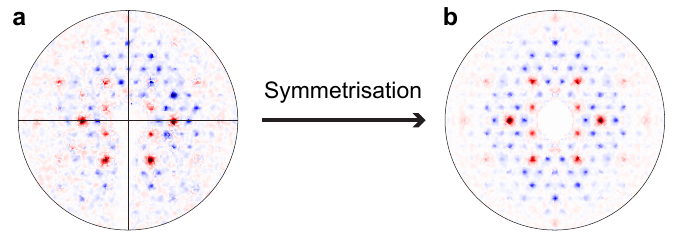}
    \caption{Difference map symmetrisation process for clearer visualization of changes. \textbf{a} Average of raw difference patterns acquired from 3--5 ps. The two mirror planes expected for this structure are superimposed. \textbf{b} Symmetrised difference map obtained by averaging signals across the mirror planes.
  }
    \label{fig:Symmetrisation}
\end{figure}

To more clearly visualise the diffraction changes, we symmetrise the difference maps displayed in the article (Figures 1-3) by averaging signals across the two mirror plane symmetries. An example of this process for the difference map averaged from 3--5 ps (Figure 3a) is shown in Supplementary Figure~\ref{fig:Symmetrisation}. This process not only enhances signal-to-noise, but also averages out some of the asymmetric changes introduced by the apparent sample buckling discussed above, making contributions from intrinsic structural change more clearly visible. We note that this process is not used to calculate quantitative signals; instead, whole pattern fitting is performed on the non-symmetrised diffraction patterns as discussed in Supplementary Note~\ref{sec:WPF} and groups of peak intensities are tracked as a function of time. 

\newpage

\section{Quantifying diffraction peak changes using whole pattern fitting}
\label{sec:WPF}

\begin{figure}[h!]
    \centering
    \includegraphics[width=13cm]{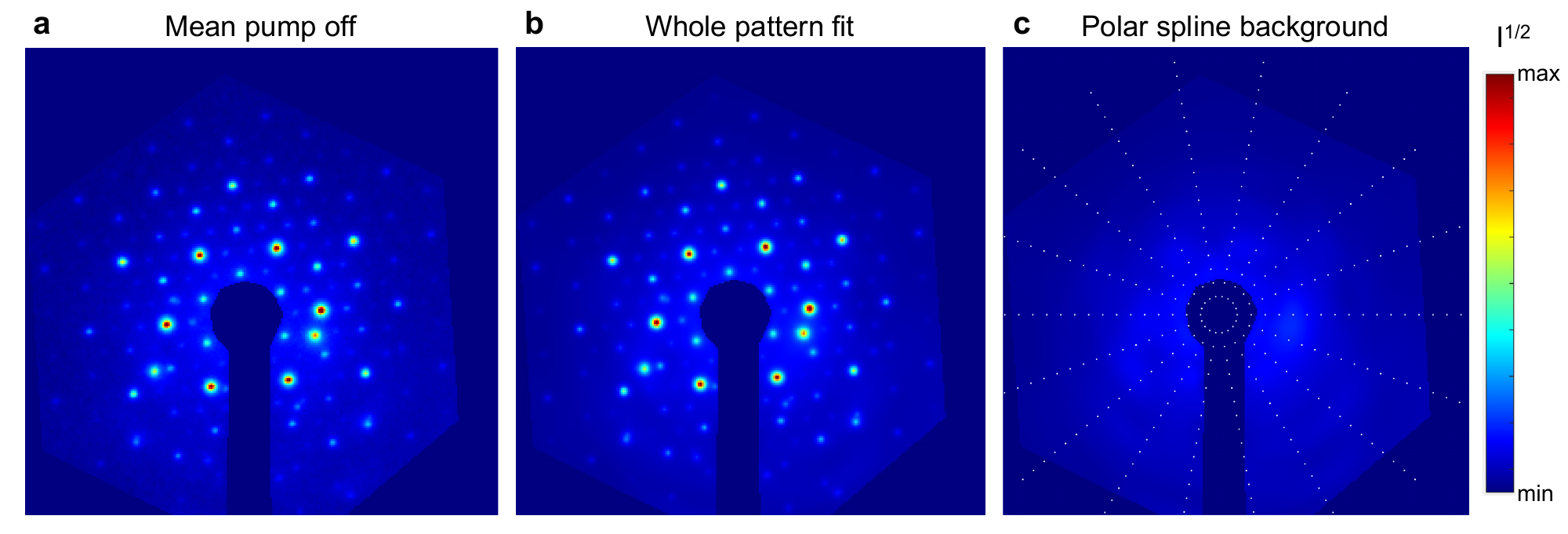}
    \caption{Whole pattern fitting of the mean pump off image. \textbf{a} Measured mean pump off image. \textbf{b} Whole pattern fit function. \textbf{c} Spline background function with knot points indicated as white dots. All images show the square root of the intensity.}
    \label{fig:WPF}
\end{figure}

To quantify diffraction peak intensities in the patterns, we performed whole pattern fitting (WPF) in which entire non-symmetrised diffraction patterns are fit with a background function and peak functions simultaneously. Compared to fitting peaks individually, WPF further constrains the fit by enforcing common features between the diffraction peaks, such as shape and positioning on a reciprocal lattice. In this case, the components include the background, the Si diffraction, the \ce{TaTe2} primary diffraction, and the \ce{TaTe2} superlattice diffraction. Within each of the three diffraction peak groups, peak positions are defined using reciprocal lattice vectors for that group and all peaks within the group are constrained to have the same shape (though varying intensities). All diffraction groups share the same pattern centre.  

To capture effectively the peak tails, the peaks are fit with generalised Gaussian functions of the form:
\begin{equation}
    I(r) = Ae^{-(r/(\sqrt{2}\alpha))^{\beta/2}}
\end{equation}
Here, r is the radial distance from the peak center, I is the intensity, A is the amplitude, $\beta$ is the shape parameter (normal distribution for $\beta$=2, stronger tails for decreasing $\beta$), and $\alpha$ is the width parameter (equivalent to the rms width when $\beta$=2).

The background fit function is a 2D spline comprised of 4th order B-splines, or basis splines, generated on a polar grid with knot points as shown in Supplementary Figure~\ref{fig:WPF}c. Such a spline has continuous 1st and 2nd derivatives everywhere on the grid, including at the knot points. The spline is fit to the data by fitting the magnitudes of the B-spline components. 

The UED patterns undergo significant preprocessing. First, the dark background reference is obtained by computing the median image of the acquired dark background frames. Then, for each time point, the dark background is subtracted, x-ray spikes are removed, frames are normalised by the total signal and aligned to 0.1 pixel precision using masked cross-correlation and Fourier shift, and then finally averaged to obtain background subtracted and processed pump ON and pump OFF images. 

To extract the time-resolved peak intensities, WPF is performed on the entire series of diffraction patterns. First, WPF is performed on the mean pump OFF image allowing all background and diffraction peak group parameters to vary. This is shown as an example in Supplementary Figure~\ref{fig:WPF}. Then, WPF is performed on the individual frames with added constraints: the overall background intensity, individual B-spline components, and the peak $\beta$ factor are fixed, whereas the individual peak intensities and peak group $\alpha$ vary. 

\newpage
\section{Transient data fitting}
\label{sec:TransientFit}

The transient data were fit using the following model function assuming exponential kinetics: 
\begin{equation}
    F(t) = H(t-t_{0})\cdot[c1\cdot(1-\exp(-\frac{t-t_{0}}{\tau}))]+C \label{eq:fiteq}
\end{equation}
where $H(t-t_{0})$ is the Heaviside step function, $\tau$ is the time constant, $t_{0}$ is time zero, $c_{1}$ and C are amplitude and constant offset, respectively. The fit function, \textit{F(t)} was numerically convolved with a Gaussian function to account for the instrument time-resolution of  $\sigma_{IRF} =  0.75$ ps. Individual fits of the data are provided in Supplementary Figure \ref{fig:CombinedFits} and the recovered values of the fit constants are collated in Supplementary Table \ref{tab:FitMainPeaks} . 

\begin{figure}[h!]
    \centering
    \includegraphics[width=15cm]{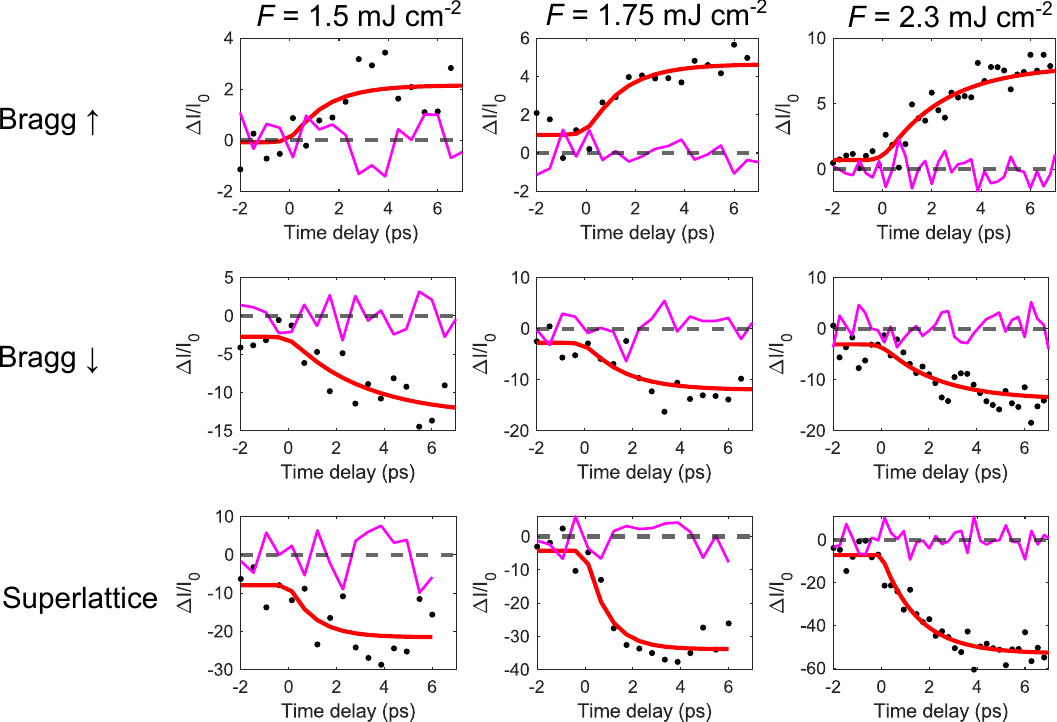}
    \caption{Transient fits of the data using the exponential kinetics in Eq.~\ref{eq:fiteq} convoluted with the 0.75~ps instrument time resolution, applied to the dynamics of Bragg$\uparrow$, Bragg$\downarrow$ and superlattice peaks for different excitation fluences of 1.5~mJ~cm$^{-2}$,  1.75~mJ~cm$^{-2}$ and  2.3~mJ~cm$^{-2}$. Fit ({\color{red} \rule[1.25 pt]{16 pt}{2 pt}}) Residual ({\color{magenta} \rule[1.25 pt]{16 pt}{2 pt}})
  }
    \label{fig:CombinedFits}
\end{figure}

\begin{table}[ht!]
\centering
\caption{Values of amplitude (A) and time constant ($\tau$) recovered from fits of time-dependent curves of Bragg$\uparrow$, Bragg$\downarrow$ and Superlattice peaks at 1.5 mJ cm$^{-2}$, 1.75 mJ cm$^{-2}$, and 2.3 mJ cm$^{-2}$.  Errors associated with the extracted values are the standard errors of the fit. }
\begin{tabular}{@{}ccccccccc@{}}
\cmidrule(lr){2-3} \cmidrule(lr){5-6} \cmidrule(l){8-9}
 & \multicolumn{2}{c}{\textbf{1.5 mJ cm$^{-2}$}} &  & \multicolumn{2}{c}{\textbf{1.75 mJ cm$^{-2}$}} &  & \multicolumn{2}{c}{\textbf{2.3 mJ cm$^{-2}$}} \\ \cmidrule(lr){2-3} \cmidrule(lr){5-6} \cmidrule(l){8-9} 
 & A & $\tau$ (ps) &  & A & $\tau$ (ps) &  & A & $\tau$ (ps) \\ \cline{2-3} \cline{5-6} \cline{8-9}
Bragg$\uparrow$ & $3.0 \pm 0.6$ & $1.4 \pm 1.6$ &  & $4.9 \pm 0.5$ & $1.4 \pm 0.7$ &  & $8.0 \pm 0.9$ & $2.4 \pm 0.8$ \\
Bragg$\downarrow$ & $-10.3 \pm 2.8$ & $3\pm 1.6$ &  & $-11.2 \pm 2.1$ & $1.6 \pm 1.3$ &  & $-13.1 \pm 2.3$ & $2.3 \pm 1.3$ \\
Superlattice & $-13.5\pm 3.4$ & $1.1 \pm 1.5$ &  & $-29.6 \pm 2.8$ & $0.9 \pm 0.4$ &  & $-46.6 \pm 2.4$ & $1.4\pm 0.3$
\label{tab:FitMainPeaks}
\end{tabular}
\end{table}

\newpage
\section{Individual diffraction order dynamics}
\begin{figure}[h!]
    \centering
    \includegraphics[width=10cm]{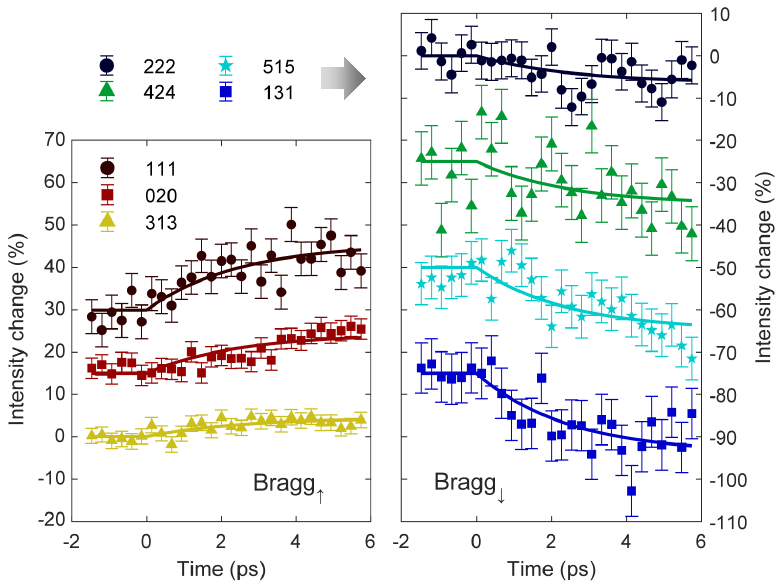}
    \caption{Dynamics of diffraction orders underlying the Bragg$\uparrow$ and Bragg$\downarrow$ curves, symmetry-averaged each from the same family of $\{hkl\}$ peaks and indexed as per the high-temperature pattern. The curves are vertically offset for clarity. Lines: scaled dynamics from Figure~2c (same time constants) as a guide to comparing the trends. Error bars in the data indicate standard error calculated using the distribution of laser-off signals compared to the mean laser-off signal over the course of the measurement.}
    \label{fig:singleDiffPlots}
\end{figure}

The Bragg$\uparrow$ and Bragg$\downarrow$ curves in Figure~2c of the main text are averages of the underlying diffraction orders. Supplementary Figure~\ref{fig:singleDiffPlots} shows the dynamics of the corresponding individual orders and evidences that the curves for each group, while exhibiting a lower signal-to-noise, display comparable dynamics as confirmed by the superimposed scaled dynamics of Figure 2c. We note that the fluctuations of the data for each curve scale with the independently-obtained error bars that correspond to $\pm 1$ standard error $\sigma_I$.

\begin{figure}[h!]
    \centering
    \includegraphics[width=9cm]{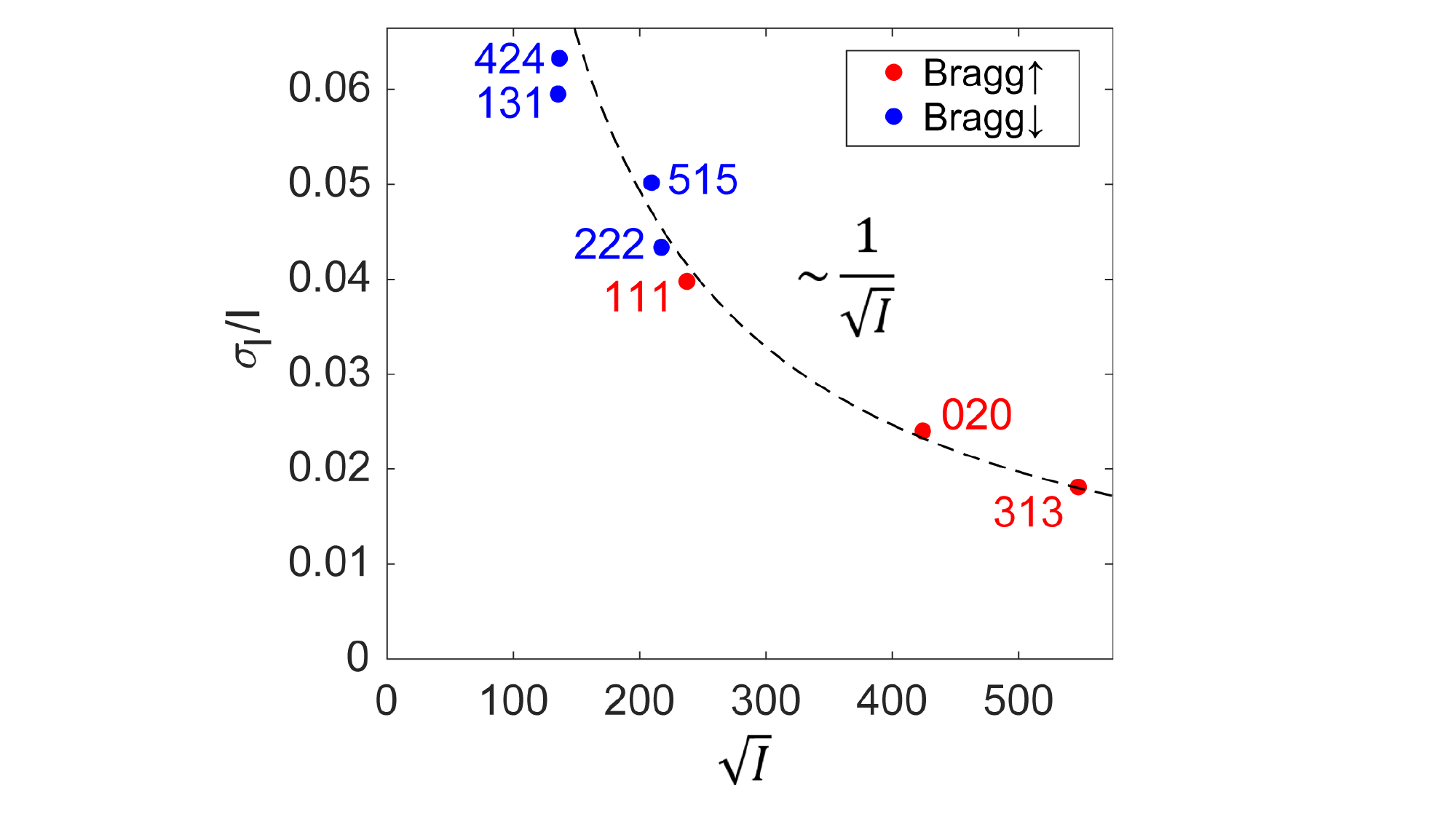}
    \caption{Scaling of the measurement error with diffraction order intensity. Relative error $\sigma_I$/$I$ versus the average $\sqrt{I}$ is shown for each diffraction order, color-coded for Bragg$\uparrow$ (red) and Bragg$\downarrow$ (blue) peaks.}
    \label{fig:singleDiffErr}
\end{figure}

As noted in the Methods section, the standard errors $\sigma_I$ are derived from the distribution of peak intensities obtained from a sequence of "pump off" images. Supplementary Figure~\ref{fig:singleDiffErr} shows the error $\sigma_I$ normalized to the peak intensity $I$ for each diffraction order. The scaling reflects a "shot noise"-like behaviour (as expected of the intensified CCD camera used in the experiments) where $\sigma_I \propto \sqrt{I}$ and thus $\sigma_I/I \propto 1/\sqrt{I}$. The lower intensities of the Bragg$\downarrow$ orders explain the higher noise in the Bragg$\downarrow$ curve as compared to the Bragg$\uparrow$ curve.

\newpage
\section{Sample optical properties from reflectivity mapping}
\label{sec:Reflectivity}

\begin{figure}[h!]
    \centering
    \includegraphics[width=15cm]{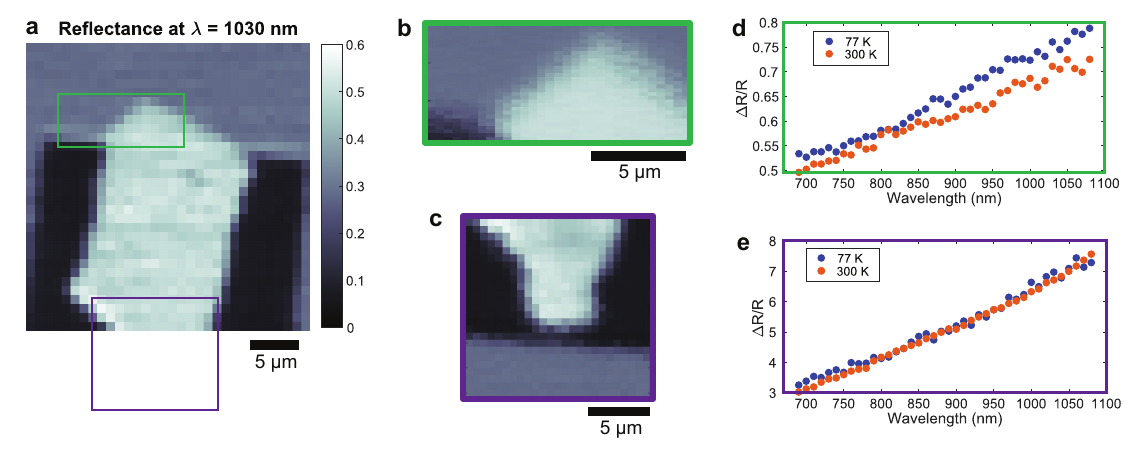}
    \caption{Reflectivity of \ce{TaTe2} sample. \textbf{a} Reflectance map across the \ce{TaTe2} flake measured at 1030 nm wavelength at 77 K. \textbf{b, c} Zoomed-in maps at 77 K of region over Si-supported \ce{Si3N4} and region over suspended \ce{Si3N4}. \textbf{d, e} Reflectance contrast spectra from region over Si-supported \ce{Si3N4} and region over suspended \ce{Si3N4}.}
    \label{fig:reflectivity}
\end{figure}

In thin films, optical propagation involves coherent interference between waves transmitted through and reflected from each interface between layers. The reflectivity, absorbance, and transmittance of a thin film sample depends not only on the thickness and dielectric function of the film, but also on that of the substrate materials. Here, we take advantage of this behaviour to extract the dielectric function of \ce{TaTe2} in the near-infrared range and estimate the absorption of the sample. We measure the reflectance in two regions of the flake over different substrates and, using the transfer-matrix method to model the reflectance in these cases, we retrieve the complex refractive index. 

We use the tmm python package\cite{byrnes2016tmm} to model the optical propagation, employing previously measured dielectric functions for \ce{Si3N4}\cite{luke2015Si3N4nk}, \ce{Si}\cite{schinke2015SiRefractiveIndex}, and \ce{SiO2}\cite{rodriguez2016SiO2nk}. In using this method, we approximate the laser as a plane wave and the in-plane dielectric function as isotropic.

Reflectance mapping was performed in a confocal laser scanning microscope. A mode-locked Ti:Sapphire laser (Coherent Chameleon) generates the incident beam, which is focused by a Nikon 40x objective lens with a numerical aperture of 0.6NA onto the surface of the sample. The reflected light was collected through the same objective lens, coupled in free space into an Andor Kymera 328i spectrometer equipped with a 50 g mm$^{-1}$, 600-nm blaze grating and detected on an Andor iDus CCD camera. The output wavelength of the laser was scanned from 690 nm to 1080 nm in steps of 10 nm, with a FWHM of about 9 nm at each wavelength, and a hyper-spectral map of the light reflected on the surface of the sample was measured at every step. The microscope’s alignment and focus were adjusted before each measurement. A continuously variable ND filter wheel was used to compensate for the power fluctuations of the laser emission as a function of wavelength.

The sample was mounted in a Janis ST-500 flow cryostat using liquid nitrogen to reach a base temperature of 77 K. We mapped the reflectivity by scanning ASI MS-2009B XY motorised translation stages supporting the sample and collecting the reflected spectrum at each position. This hyperspectral measurement was controlled using the open-source ScopeFoundry software platform\cite{durhamScopeFoundry,ScopeFoundryWebsite}.

To calibrate the reflectance, we measured the reflected signal from a 100 nm thick gold film and the \ce{Si3N4}-covered Si TEM chip under the same conditions. For instance at 1030 nm, after correcting for the reflectivity of gold (99\%), we obtained a reflectance from the Si chip of about 28\%. Using this and the measured reflectance contrast, $\frac{\Delta R}{R}$, as a function of laser wavelength between freestanding \ce{Si3N4} and the Si chip, we fit the thickness of the membrane and the Si native oxide, finding a best fit for 28 nm of \ce{Si3N4} and a 2 nm \ce{SiO2} native oxide layer on the Si chip. 

Reflectance maps at 77 K for 1030 nm light are shown in Supplementary Figure~\ref{fig:reflectivity}a-c. We observe that the reflectivity is largely homogeneous across the flake, with deviations mainly at topographic features such as flake edges and small pieces of additional \ce{TaTe2} sitting on top from the sample transfer process (see AFM map in Supplementary Note~\ref{sec:Sample}). This is evidence that the sample did not experience damage due to optical laser beam during the UED experiments. We also observe similar reflectivity between regions over the Si chip and the suspended \ce{Si3N4} membrane. 

From the zoomed-in maps taken at each wavelength, we compute the reflectance contrast spectra between \ce{TaTe2} and the two substrate types: the Si chip and the suspended \ce{Si3N4} membrane (Supplementary Figure~\ref{fig:reflectivity}d,e). We observe similar reflectance values and trends with the wavelength at 77 K and 300 K, though slightly higher reflectance is observed in the region over the Si chip at 77 K which may be due to formation of the LT phase.

\begin{figure}[h!]
    \centering
    \includegraphics[width=15cm]{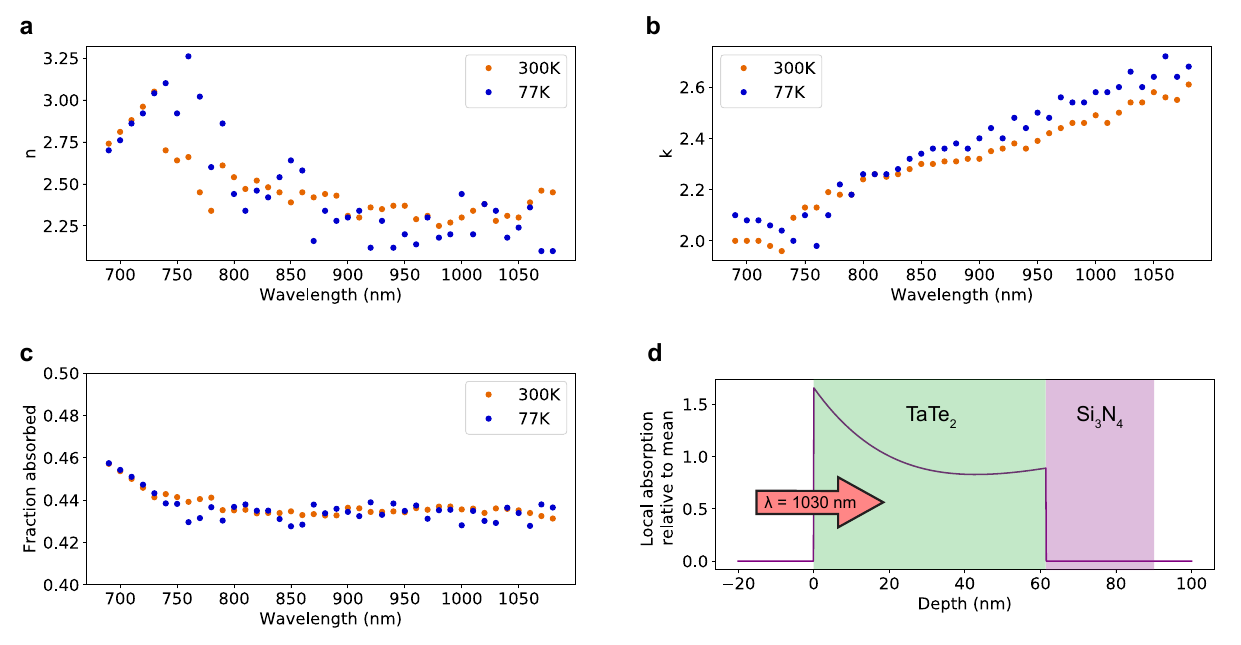}
    \caption{Dielectric function and absorption spectra extracted using the transfer-matrix method. \textbf{a} Best-fit real part of the refractive index of \ce{TaTe2}. \textbf{b} Best-fit imaginary part of the refractive index of \ce{TaTe2}. \textbf{c} Calculated absorption of the \ce{TaTe2} flake supported by \ce{Si3N4}. \textbf{d} Calculated local absorption as a function of depth.}
    \label{fig:absorbance}
\end{figure}

\begin{figure}[h!]
    \centering
    \includegraphics[width=15cm]{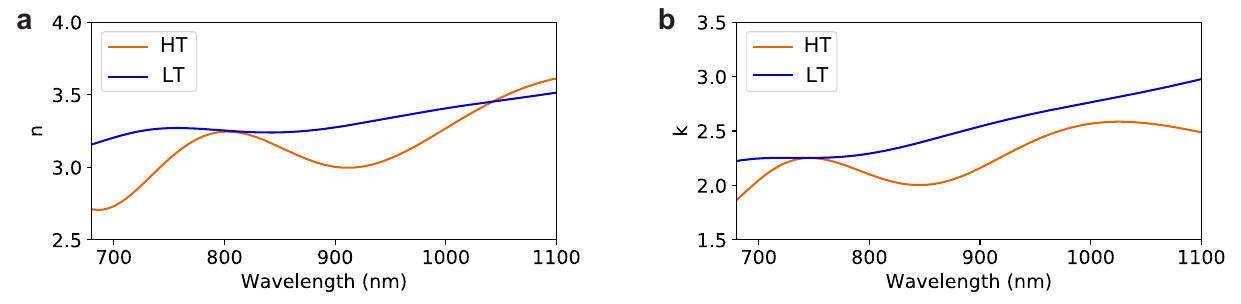}
    \caption{DFT-calculated refractive index for light polarised along the b axis. \textbf{a} Real part for LT and HT phases. \textbf{b} Imaginary part for LT and HT phases.}
    \label{fig:simnk}
\end{figure}

Using the transfer-matrix method, we optimise the real and imaginary parts of the refractive index to match the reflectance contrast in both regions, shown in Figure~\ref{fig:absorbance}a,b. The extracted refractive index shows smooth and largely monotonic behaviour in this region, and it is similar at both temperatures. From the refractive index, \textit{n} we calculated the total absorption spectrum via the transfer-matrix method of the \ce{Si3N4}-supported \ce{TaTe2} sample (Figure~\ref{fig:absorbance}c). At 1030 nm wavelength, the fraction absorbed is about 44\% at both 77 K and 300 K. We also compute the absorption profile over the depth of the film at 1030 nm, shown in Supplementary Figure~\ref{fig:absorbance}d. As the absorption depth is 33 nm, we calculate significant variation in absorption through the 60 nm thickness of the \ce{TaTe2} film: Material near the top surface absorbs more than 1.5 times more photons than average through the depth.

To check whether the obtained refractive index behaviour is reasonable for this system, we calculate frequency dependent dielectric response of \ce{TaTe2} using VASP under independent particle approximation. The calculated refractive indices of the relaxed LT and HT structures of \ce{TaTe2} for light polarised along the crystal\textit{ b} axis is shown in the range of our measurements in Supplementary Figure~\ref{fig:simnk}. Though direct quantitative comparison to the experiment is complicated by experimental factors such as sample tilt, laser polarisation character, laser bandwidth, and angular spread of the laser beam, we make a few qualitative observations. First, the dielectric function of the LT and HT phases have similar magnitude over this wavelength range, both in the calculation and in experiment. Second, the imaginary component corresponding to absorptivity shows good agreement between simulation and experiment: the magnitude is between 2.0 and 2.7 and generally increases for longer wavelengths. Thirdly, the LT phase appears to have higher k over most of the wavelength range both in measurement and experiment: This is consistent with the increased density of states around the Fermi level in the LT phase observed in DFT calculations (see Supplementary Note~\ref{sec:DFT}).

\newpage

\section{Thermalisation and fluence dependent dynamics}
\label{sec:FluenceDep}

\begin{figure}[h!]
    \centering
    \includegraphics[width=13cm]{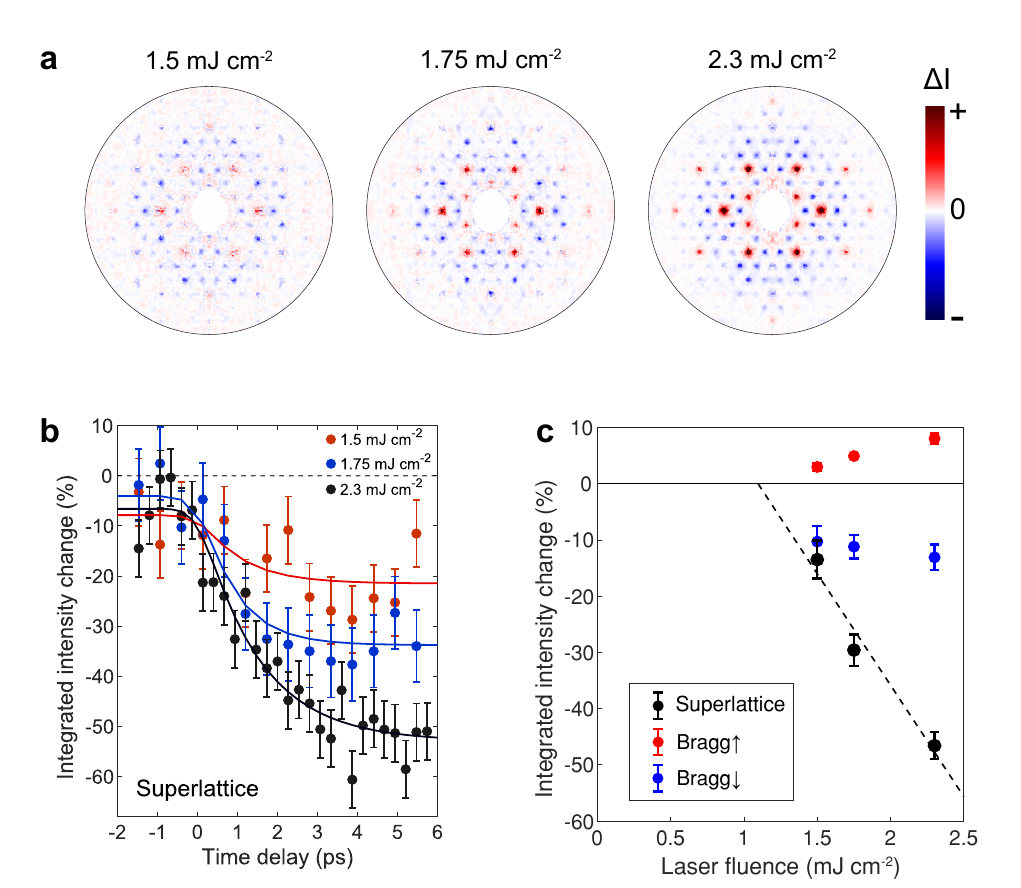}
    \caption{Photoinduced superstructure melting at varying laser fluence. \textbf{a} Representative difference images for early (centred around $\approx\!~4$~ps) time delays for a laser fluence of 1.5 mJ cm$^{-2}$, 1.75 mJ cm$^{-2}$, and 2.3 mJ cm$^{-2}$. For improved signal-to-noise these are calculated by averaging difference patterns over a 3--5~ps time range. \textbf{b} Initial superlattice peak dynamics for varying laser fluence. Exponential fits are superimposed as solid lines, with time constants of $1.07\pm0.92$ ps, $0.9\pm0.32$ ps, and $1.44\pm0.27$ ps for the three fluences in increasing order. Error bars indicate standard error calculated using the distribution of laser-off signals compared to the mean laser-off signal over the course of the measurement. \textbf{c} Magnitude of peak subset changes for varying laser fluence. These are the difference between averaged values over -5 to 0 ps and 4 to 6 ps range, and error bars indicate standard error, calculated from those for the individual points. A linear fit to the superlattice peak intensities is shown as a black dashed line. }
    \label{fig:FluenceDep}
\end{figure}

In addition to the experiment at 2.3 mJ cm$^{-2}$ highlighted in the main text, we also measured the initial ultrafast optical melting dynamics at 1.5 mJ cm$^{-2}$ and 1.75 mJ cm$^{-2}$. Difference patterns averaged over a 3--5~ps time range are shown in Supplementary Figure~\ref{fig:FluenceDep}a. The magnitude of observed changes increases with the incident fluence over this range, and the mixture of positive and negative changes at all fluences resembles that observed for the thermal phase change shown in Figure 1. The quantitative temporal profiles of the superlattice peak intensities are shown in Supplementary Figure~\ref{fig:FluenceDep}b. The measured time constants of PLD suppression fall within the range of 1 to 1.5 ps for all fluences. The increasing magnitude of superlattice peak reduction, as well as of change in $\text{Bragg}\uparrow$ and $\text{Bragg}\downarrow$, is shown in Supplementary Figure~\ref{fig:FluenceDep}c. Moreover, as the amplitude of the PLD superlattice peaks tracks the order parameter, the intensity of these peaks are mainly sensitive to the structural change and impacted less by thermal effects. Thus, these peaks are most suited to extract the threshold fluence for the structural phase transition. A linear fit on the PLD peak intensities versus fluence (dashed line, Fig.~\ref{fig:FluenceDep}c) suggests a zero-intercept around $\approx\!1.1$~mJ cm$^{-2}$.

\newpage
\noindent We can estimate the final temperature of the sample in our UED experiments after complete thermalisation following laser absorption at 1030 nm. In this, it is assumed that 44\% of the incident light gets absorbed (based on optical measurements in Supplementary Note~\ref{sec:Reflectivity}) in the film of 61.5 nm thickness (as determined in Supplementary Note~\ref{sec:Sample}). We calculated the corresponding temperature increase using the temperature-dependent heat capacity $C_{p}(T)$ from ref.~\citenum{Sorgel2006}, by taking into account the material's density\cite{Sorgel2006} $\rho$ = 9.3 g cm$^{-3}$ and molar mass M = 436.15 g mol$^{-1}$. This resulted in thermally equilibrated temperatures of \noindent$T_{therm}$ = 125 K, 130~K and 177~K for the three fluences, respectively, of $F = 1.5$~mJ~cm~$^{-2}$, 1.75~mJ~cm~$^{-2}$, and 2.3~mJ~cm~$^{-2}$.

\begin{figure}[h!]
    \centering
    \includegraphics[width=7cm]{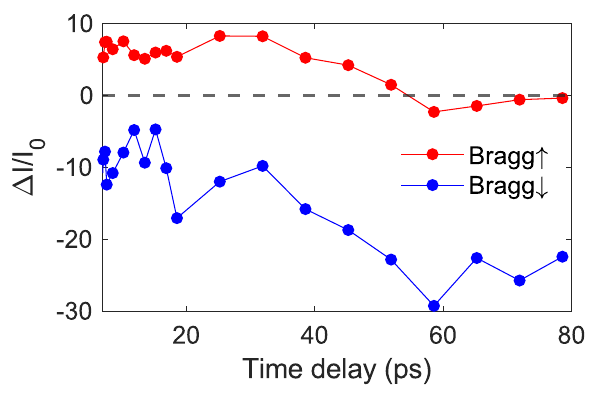}
    \caption{Long term Bragg peak dynamics. Evolution of Bragg($\uparrow$) and Bragg($\downarrow$) peaks over the course of scan. }
    \label{fig:BraggLongTerm}
\end{figure}

Experimentally, the lattice thermalisation into a quasi-equilibrium temperature is observed on extended time scales via a global decrease of the primary lattice peak intensities. Supplementary Fig.~\ref{fig:BraggLongTerm} shows the corresponding dynamics of the Bragg peaks at long time delays, for a fluence of 2.3 mJ cm$^{-2}$. 

\begin{figure}[h!]
    \centering
    \includegraphics[width=7cm]{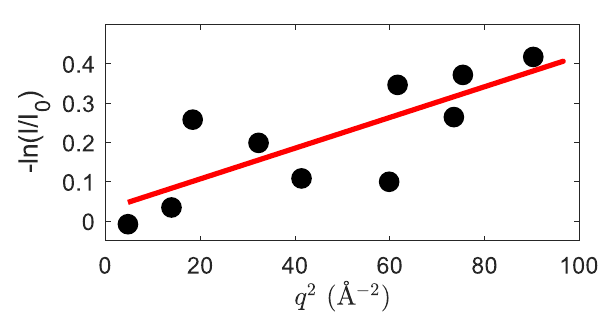}
    \caption{Log graph of relative Bragg peak intensity of the photo-excited TaTe$_2$ vs the square momentum transfer $q^2$ of selected peaks. The data was taken at 2.3 mJ cm$^{-2}$ fluence and evaluated at the longest time delay of 78~ps. The fit corresponds to $\langle u^2\rangle \approx\! 0.012$~\AA$^2$.}
    \label{fig:Debye_Waller_Fit}
\end{figure}

To gain insight into the conditions we can estimate the Debye-Waller factor in the photoexcited state relative to the 10 K ground state. Supplementary Fig.~\ref{fig:Debye_Waller_Fit} plots the logarithm of the relative Bragg peak intensity (normalised to the intensity before excitation) versus momentum. The linear fit corresponds to an induced mean square displacement $\langle u_{\rm ind}^2\rangle \approx\!0.012\pm 0.004$~\AA$^2$. This was evaluated at the fluence of 2.3~mJ cm$^-{2}$ and longest delay of 78~ps, where a thermalised lattice is expected. From this, we obtain a Debye-Waller factor $B = 8\pi^2/3\langle u_{\rm in}^2\rangle \approx\!0.3$~\AA$^2$. We can compare this value with an equilibrium measurement\cite{Sorgel2006} which obtained a displacement parameter $\langle u^2\rangle~\approx\!0.014$~\AA$^2$ at 150~K from X-ray measurements of TaTe$_2$. These values compare well, and are in line with the expected temperature being below $T_{\rm c}$ at long time delays. However, we caution that multiple scattering effects are not taken into account in this analysis.

\newpage
\section{Sample reversibility}

In pump-probe experiments, sample damage could result from absorption of very intense laser pulses and from knock-on damage due to the high energy electrons. The latter depends on the dose defined as the number of particles per unit area, which if exceeded beyond a critical threshold can lead to formation of reactive species, e.g. radicals, create vacancies and other related defects. However, in UED this usually does not pose an issue due to the relatively large focus diameters of the electron beam ($>100~\mu$m). In our study, we estimate a dose of about 0.05 electrons per nm$^{2}$ per pulse, which is too low to cause damage.

\begin{figure}[h!]
    \centering
    \includegraphics[width=8cm]{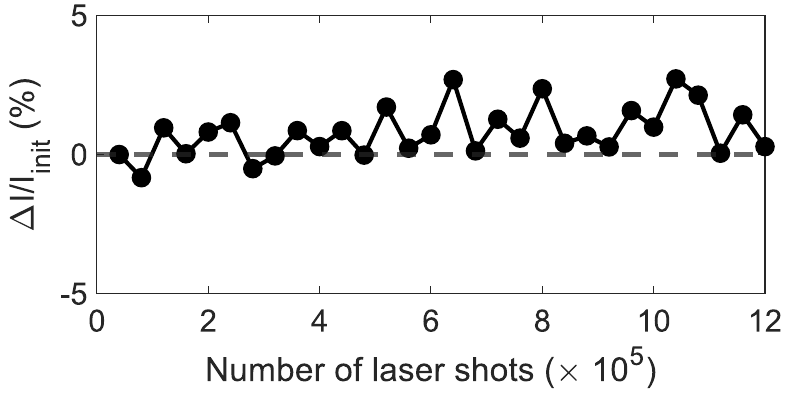}
    \caption{Change of ground-state peak intensity during the experiment $\Delta I/I_{\rm init} = \left(I(n)-I_{\rm init}\right )/I_{\rm init}$, plotted as a function of the total accumulated laser shots. The fluctuations remain within the range expected from electron beam drifts.}
    \label{fig:Reversibility}
\end{figure}

\begin{figure}[h!]
    \centering
    \includegraphics[width=14cm]{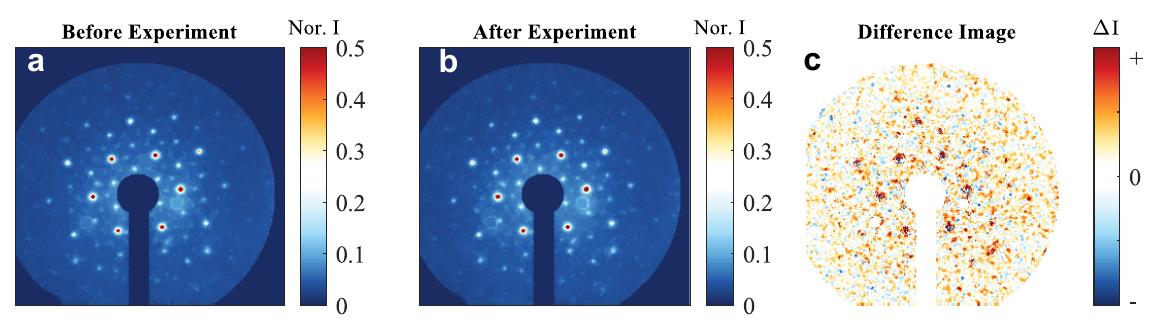}
        \caption{Diffraction images of \ce{TaTe2} from experiment at 2.3 mJ cm$^{-2}$. \textbf{a} Measured pattern before laser exposure. \textbf{b} Measured pattern after the experiment, i.e. exposure to $>10^6$ laser shots. \textbf{c} Difference image between before and after experiment showing only minor system drifts.}
    \label{fig:DamageImages}
\end{figure}

Under the 2.3 mJ cm$^{-2}$ near-IR pump, UED probe conditions we did not observe any sample damage over the extent of the UED measurements as judged by several diagnostics. First, no damage was visible macroscopically, confirmed by homogeneous near-IR reflectance (cf. Supplementary Fig~\ref{fig:reflectivity}a) and AFM measurements. Moreover, we confirmed that the sample maintained its crystallinity during the experiments by measuring the peak intensity $I(n)$ after each laser and electron exposure sequence containing about $4\times10^4$ shots and comparing that to the intensity $I_{\rm init}$ of the peak before the experiment as:

\begin{equation}
    \frac{\Delta I}{I_{\rm init}} = \frac{\left(I(n)-I_{\rm init}\right )}{I_{\rm init}}
\end{equation}

\noindent where $I(n)$ is intensity of the peak after each laser on sequence and $I_{\rm init}$ is the intensity of the peak before laser exposure. Supplementary Fig.~\ref{fig:Reversibility} shows the intensities after several thousand laser shots, which remained within the expected $\approx$~1--2\% level of long-term experimental drifts. This, and the absence of any reduction in intensity, confirms that the sample remained intact even after the extended exposure.  This is further supported by comparing the normalised diffraction patterns taken before and after the experiment (Supplementary Fig.~\ref{fig:DamageImages}). The two patterns are almost indistinguishable except for noise and a few-pixel shift of the overall patterns due to slight beam-pointing changes.

\begin{thebibliography}{0}%
\makeatletter
\providecommand \@ifxundefined [1]{%
 \@ifx{#1\undefined}
}%
\providecommand \@ifnum [1]{%
 \ifnum #1\expandafter \@firstoftwo
 \else \expandafter \@secondoftwo
 \fi
}%
\providecommand \@ifx [1]{%
 \ifx #1\expandafter \@firstoftwo
 \else \expandafter \@secondoftwo
 \fi
}%
\providecommand \natexlab [1]{#1}%
\providecommand \enquote  [1]{``#1''}%
\providecommand \bibnamefont  [1]{#1}%
\providecommand \bibfnamefont [1]{#1}%
\providecommand \citenamefont [1]{#1}%
\providecommand \href@noop [0]{\@secondoftwo}%
\providecommand \href [0]{\begingroup \@sanitize@url \@href}%
\providecommand \@href[1]{\@@startlink{#1}\@@href}%
\providecommand \@@href[1]{\endgroup#1\@@endlink}%
\providecommand \@sanitize@url [0]{\catcode `\\12\catcode `\$12\catcode
  `\&12\catcode `\#12\catcode `\^12\catcode `\_12\catcode `\%12\relax}%
\providecommand \@@startlink[1]{}%
\providecommand \@@endlink[0]{}%
\providecommand \url  [0]{\begingroup\@sanitize@url \@url }%
\providecommand \@url [1]{\endgroup\@href {#1}{\urlprefix }}%
\providecommand \urlprefix  [0]{URL }%
\providecommand \Eprint [0]{\href }%
\providecommand \doibase [0]{http://dx.doi.org/}%
\providecommand \selectlanguage [0]{\@gobble}%
\providecommand \bibinfo  [0]{\@secondoftwo}%
\providecommand \bibfield  [0]{\@secondoftwo}%
\providecommand \translation [1]{[#1]}%
\providecommand \BibitemOpen [0]{}%
\providecommand \bibitemStop [0]{}%
\providecommand \bibitemNoStop [0]{.\EOS\space}%
\providecommand \EOS [0]{\spacefactor3000\relax}%
\providecommand \BibitemShut  [1]{\csname bibitem#1\endcsname}%
\let\auto@bib@innerbib\@empty
%</preamble>
\end{thebibliography}%


\begin{thebibliography}{10}
\expandafter\ifx\csname url\endcsname\relax
  \def\url#1{\texttt{#1}}\fi
\expandafter\ifx\csname urlprefix\endcsname\relax\def\urlprefix{URL }\fi
\providecommand{\bibinfo}[2]{#2}
\providecommand{\eprint}[2][]{\url{#2}}

\bibitem{Tok17}
\bibinfo{author}{Tokura, Y.}, \bibinfo{author}{Kawasaki, M.} \&
  \bibinfo{author}{Nagaosa, N.}
\newblock \bibinfo{title}{Emergent functions of quantum materials}.
\newblock \emph{\bibinfo{journal}{Nat. Phys.}} \textbf{\bibinfo{volume}{13}},
  \bibinfo{pages}{1056--1068} (\bibinfo{year}{2017}).

\bibitem{Kei15}
\bibinfo{author}{Keimer, B.}, \bibinfo{author}{Kivelson, S.~A.},
  \bibinfo{author}{Norman, M.~R.}, \bibinfo{author}{Uchida, S.} \&
  \bibinfo{author}{Zaanen, J.}
\newblock \bibinfo{title}{From quantum matter to high-temperature
  superconductivity in copper oxides}.
\newblock \emph{\bibinfo{journal}{Nature}} \textbf{\bibinfo{volume}{518}},
  \bibinfo{pages}{179--86} (\bibinfo{year}{2015}).

\bibitem{Has10}
\bibinfo{author}{Hasan, M.~Z.} \& \bibinfo{author}{Kane, C.~L.}
\newblock \bibinfo{title}{Colloquium: Topological insulators}.
\newblock \emph{\bibinfo{journal}{Rev. Mod. Phys.}}
  \textbf{\bibinfo{volume}{82}}, \bibinfo{pages}{3045} (\bibinfo{year}{2010}).

\bibitem{Arm18}
\bibinfo{author}{Armitage, N.~P.}, \bibinfo{author}{Mele, E.~J.} \&
  \bibinfo{author}{Vishwanath, A.}
\newblock \bibinfo{title}{Weyl and {Dirac} semimetals in three-dimensional
  solids}.
\newblock \emph{\bibinfo{journal}{Rev. Mod. Phys.}}
  \textbf{\bibinfo{volume}{90}}, \bibinfo{pages}{015001}
  (\bibinfo{year}{2018}).

\bibitem{Nov16}
\bibinfo{author}{Novoselov, K.~S.}, \bibinfo{author}{Mishchenko, A.},
  \bibinfo{author}{Carvalho, A.} \& \bibinfo{author}{Castro~Neto, A.~H.}
\newblock \bibinfo{title}{{2D} materials and van der {Waals} heterostructures}.
\newblock \emph{\bibinfo{journal}{Science}} \textbf{\bibinfo{volume}{353}},
  \bibinfo{pages}{aac9439} (\bibinfo{year}{2016}).

\bibitem{Zha14}
\bibinfo{author}{Zhang, J.} \& \bibinfo{author}{Averitt, R.}
\newblock \bibinfo{title}{Dynamics and control in complex transition metal
  oxides}.
\newblock \emph{\bibinfo{journal}{Ann. Rev. Mat. Res.}}
  \textbf{\bibinfo{volume}{44}}, \bibinfo{pages}{19--43}
  (\bibinfo{year}{2014}).

\bibitem{Bas17}
\bibinfo{author}{Basov, D.~N.}, \bibinfo{author}{Averitt, R.~D.} \&
  \bibinfo{author}{Hsieh, D.}
\newblock \bibinfo{title}{Towards properties on demand in quantum materials}.
\newblock \emph{\bibinfo{journal}{Nat. Mater.}} \textbf{\bibinfo{volume}{16}},
  \bibinfo{pages}{1077} (\bibinfo{year}{2017}).

\bibitem{Cav01}
\bibinfo{author}{Cavalleri, A.} \emph{et~al.}
\newblock \bibinfo{title}{Femtosecond structural dynamics in \ce{VO2} during an
  ultrafast solid-solid phase transition}.
\newblock \emph{\bibinfo{journal}{Phys. Rev. Lett.}}
  \textbf{\bibinfo{volume}{87}}, \bibinfo{pages}{237401}
  (\bibinfo{year}{2001}).

\bibitem{Lee12}
\bibinfo{author}{Lee, W.~S.} \emph{et~al.}
\newblock \bibinfo{title}{Phase fluctuations and the absence of topological
  defects in a photo-excited charge-ordered nickelate}.
\newblock \emph{\bibinfo{journal}{Nature Commun.}}
  \textbf{\bibinfo{volume}{3}}, \bibinfo{pages}{838} (\bibinfo{year}{2012}).

\bibitem{Tri13}
\bibinfo{author}{Trigo, M.} \emph{et~al.}
\newblock \bibinfo{title}{Fourier-transform inelastic x-ray scattering from
  time- and momentum-dependent phonon–phonon correlations}.
\newblock \emph{\bibinfo{journal}{Nat. Phys.}} \textbf{\bibinfo{volume}{9}},
  \bibinfo{pages}{790--794} (\bibinfo{year}{2013}).

\bibitem{Otto_2018}
\bibinfo{author}{Otto, M.~R.} \emph{et~al.}
\newblock \bibinfo{title}{How optical excitation controls the structure and
  properties of vanadium dioxide}.
\newblock \emph{\bibinfo{journal}{Proc. Natl. Acad. Sci.}}
  \textbf{\bibinfo{volume}{116}}, \bibinfo{pages}{450--455}
  (\bibinfo{year}{2018}).

\bibitem{Sie19}
\bibinfo{author}{Sie, E.~J.} \emph{et~al.}
\newblock \bibinfo{title}{An ultrafast symmetry switch in a {Weyl} semimetal}.
\newblock \emph{\bibinfo{journal}{Nature}} \textbf{\bibinfo{volume}{565}},
  \bibinfo{pages}{61--66} (\bibinfo{year}{2019}).

\bibitem{Zon19}
\bibinfo{author}{Zong, A.} \emph{et~al.}
\newblock \bibinfo{title}{Evidence for topological defects in a photoinduced
  phase transition}.
\newblock \emph{\bibinfo{journal}{Nat. Phys.}} \textbf{\bibinfo{volume}{15}},
  \bibinfo{pages}{27--31} (\bibinfo{year}{2019}).

\bibitem{Rin07}
\bibinfo{author}{Rini, M.} \emph{et~al.}
\newblock \bibinfo{title}{Control of the electronic phase of a manganite by
  mode-selective vibrational excitation}.
\newblock \emph{\bibinfo{journal}{Nature}} \textbf{\bibinfo{volume}{449}},
  \bibinfo{pages}{72--74} (\bibinfo{year}{2007}).

\bibitem{Por14}
\bibinfo{author}{Porer, M.} \emph{et~al.}
\newblock \bibinfo{title}{Non-thermal separation of electronic and structural
  orders in a persisting charge density wave}.
\newblock \emph{\bibinfo{journal}{Nat. Mater.}} \textbf{\bibinfo{volume}{13}},
  \bibinfo{pages}{857--861} (\bibinfo{year}{2014}).

\bibitem{Cos17}
\bibinfo{author}{Coslovich, G.} \emph{et~al.}
\newblock \bibinfo{title}{Ultrafast dynamics of vibrational symmetry breaking
  in a charge-ordered nickelate}.
\newblock \emph{\bibinfo{journal}{Sci. Adv.}} \textbf{\bibinfo{volume}{3}},
  \bibinfo{pages}{1600735} (\bibinfo{year}{2017}).

\bibitem{Per06}
\bibinfo{author}{Perfetti, L.} \emph{et~al.}
\newblock \bibinfo{title}{Time evolution of the electronic structure of
  \ce{1T-TaS2} through the insulator-metal transition}.
\newblock \emph{\bibinfo{journal}{Phys. Rev. Lett.}}
  \textbf{\bibinfo{volume}{97}}, \bibinfo{pages}{067402}
  (\bibinfo{year}{2006}).

\bibitem{Roh11}
\bibinfo{author}{Rohwer, T.} \emph{et~al.}
\newblock \bibinfo{title}{Collapse of long-range charge order tracked by
  time-resolved photoemission at high momenta}.
\newblock \emph{\bibinfo{journal}{Nature}} \textbf{\bibinfo{volume}{471}},
  \bibinfo{pages}{490--493} (\bibinfo{year}{2011}).

\bibitem{Sob20}
\bibinfo{author}{Sobota, J.~A.}, \bibinfo{author}{He, Y.} \&
  \bibinfo{author}{Shen, Z.-X.}
\newblock \bibinfo{title}{Angle-resolved photoemission studies of quantum
  materials}.
\newblock \emph{\bibinfo{journal}{Reviews of Modern Physics}}
  \textbf{\bibinfo{volume}{93}}, \bibinfo{pages}{025006}
  (\bibinfo{year}{2021}).

\bibitem{Sip08}
\bibinfo{author}{Sipos, B.} \emph{et~al.}
\newblock \bibinfo{title}{From {M}ott state to superconductivity in
  \ce{1T-TaS2}}.
\newblock \emph{\bibinfo{journal}{Nat. Mater.}} \textbf{\bibinfo{volume}{7}},
  \bibinfo{pages}{960--5} (\bibinfo{year}{2008}).

\bibitem{Cho16}
\bibinfo{author}{Cho, D.} \emph{et~al.}
\newblock \bibinfo{title}{Nanoscale manipulation of the {M}ott insulating state
  coupled to charge order in \ce{1T-TaS2}}.
\newblock \emph{\bibinfo{journal}{Nat. Commun.}} \textbf{\bibinfo{volume}{7}},
  \bibinfo{pages}{10453} (\bibinfo{year}{2016}).

\bibitem{Miller2018}
\bibinfo{author}{Miller, D.~C.}, \bibinfo{author}{Mahanti, S.~D.} \&
  \bibinfo{author}{Duxbury, P.~M.}
\newblock \bibinfo{title}{Charge density wave states in tantalum
  dichalcogenides}.
\newblock \emph{\bibinfo{journal}{Phys. Rev. B}} \textbf{\bibinfo{volume}{97}},
  \bibinfo{pages}{045133} (\bibinfo{year}{2018}).

\bibitem{Eichberger_2010}
\bibinfo{author}{Eichberger, M.} \emph{et~al.}
\newblock \bibinfo{title}{Snapshots of cooperative atomic motions in the
  optical suppression of charge density waves}.
\newblock \emph{\bibinfo{journal}{Nature}} \textbf{\bibinfo{volume}{468}},
  \bibinfo{pages}{799--802} (\bibinfo{year}{2010}).

\bibitem{Haupt2016}
\bibinfo{author}{Haupt, K.} \emph{et~al.}
\newblock \bibinfo{title}{Ultrafast metamorphosis of a complex charge-density
  wave}.
\newblock \emph{\bibinfo{journal}{Phys. Rev. Lett.}}
  \textbf{\bibinfo{volume}{116}}, \bibinfo{pages}{016402}
  (\bibinfo{year}{2016}).

\bibitem{Vogelgesang_2017}
\bibinfo{author}{Vogelgesang, S.} \emph{et~al.}
\newblock \bibinfo{title}{Phase ordering of charge density waves traced by
  ultrafast low-energy electron diffraction}.
\newblock \emph{\bibinfo{journal}{Nat. Phys.}} \textbf{\bibinfo{volume}{14}},
  \bibinfo{pages}{184--190} (\bibinfo{year}{2017}).

\bibitem{Sto14}
\bibinfo{author}{Stojchevska, L.} \emph{et~al.}
\newblock \bibinfo{title}{Ultrafast switching to a stable hidden quantum state
  in an electronic crystal}.
\newblock \emph{\bibinfo{journal}{Science}} \textbf{\bibinfo{volume}{344}},
  \bibinfo{pages}{177} (\bibinfo{year}{2014}).

\bibitem{CYRuan2015TaS2}
\bibinfo{author}{Han, T.-R.~T.} \emph{et~al.}
\newblock \bibinfo{title}{Exploration of metastability and hidden phases in
  correlated electron crystals visualized by femtosecond optical doping and
  electron crystallography}.
\newblock \emph{\bibinfo{journal}{Sci. Adv.}} \textbf{\bibinfo{volume}{1}},
  \bibinfo{pages}{e1400173} (\bibinfo{year}{2015}).

\bibitem{Luo_2015}
\bibinfo{author}{Luo, H.} \emph{et~al.}
\newblock \bibinfo{title}{Polytypism, polymorphism, and superconductivity in
  \ce{TaSe2}$_{-x}$\ce{Te}$_{x}$}.
\newblock \emph{\bibinfo{journal}{Proc. Natl. Acad. Sci.}}
  \textbf{\bibinfo{volume}{112}}, \bibinfo{pages}{E1174--E1180}
  (\bibinfo{year}{2015}).

\bibitem{Sun2015}
\bibinfo{author}{Sun, S.} \emph{et~al.}
\newblock \bibinfo{title}{Direct observation of an optically induced charge
  density wave transition in \ce{1T-TaSe2}}.
\newblock \emph{\bibinfo{journal}{Phys. Rev. B}} \textbf{\bibinfo{volume}{92}},
  \bibinfo{pages}{224303} (\bibinfo{year}{2015}).

\bibitem{Linlin2017}
\bibinfo{author}{Wei, L.} \emph{et~al.}
\newblock \bibinfo{title}{Dynamic diffraction effects and coherent breathing
  oscillations in ultrafast electron diffraction in layered {1T-TaSeTe}}.
\newblock \emph{\bibinfo{journal}{Struct. Dyn.}} \textbf{\bibinfo{volume}{4}},
  \bibinfo{pages}{044012} (\bibinfo{year}{2017}).

\bibitem{li2019ultrafast}
\bibinfo{author}{Li, J.} \emph{et~al.}
\newblock \bibinfo{title}{Ultrafast decoupling of atomic sublattices in a
  charge-density-wave material}.
\newblock \emph{\bibinfo{journal}{ArXiv}} \bibinfo{pages}{Preprint at
  {\url{https://arxiv.org/abs/1903.09911}}} (\bibinfo{year}{2019}).

\bibitem{Wil69}
\bibinfo{author}{Wilson, J.} \& \bibinfo{author}{Yoffe, A.}
\newblock \bibinfo{title}{The transition metal dichalcogenides discussion and
  interpretation of the observed optical, electrical and structural
  properties}.
\newblock \emph{\bibinfo{journal}{Adv. Phys.}} \textbf{\bibinfo{volume}{18}},
  \bibinfo{pages}{193--335} (\bibinfo{year}{1969}).

\bibitem{Sorgel2006}
\bibinfo{author}{Sörgel, T.}, \bibinfo{author}{Nuss, J.},
  \bibinfo{author}{Wedig, U.}, \bibinfo{author}{Kremer, R.} \&
  \bibinfo{author}{Jansen, M.}
\newblock \bibinfo{title}{A new low temperature modification of
  \ce{TaTe2}{\textemdash}comparison to the room temperature and the
  hypothetical \ce{1T-TaTe2} modification}.
\newblock \emph{\bibinfo{journal}{Mater. Res. Bull.}}
  \textbf{\bibinfo{volume}{41}}, \bibinfo{pages}{987 -- 1000}
  (\bibinfo{year}{2006}).

\bibitem{Doublet_2000}
\bibinfo{author}{Doublet, M.-L.}, \bibinfo{author}{Remy, S.} \&
  \bibinfo{author}{Lemoigno, F.}
\newblock \bibinfo{title}{Density functional theory analysis of the local
  chemical bonds in the periodic tantalum dichalcogenides \ce{TaX2} ({X = S,
  Se, Te})}.
\newblock \emph{\bibinfo{journal}{J. Chem. Phys.}}
  \textbf{\bibinfo{volume}{113}}, \bibinfo{pages}{5879--5890}
  (\bibinfo{year}{2000}).

\bibitem{Gao2018}
\bibinfo{author}{Gao, J.~J.} \emph{et~al.}
\newblock \bibinfo{title}{Origin of the structural phase transition in
  single-crystal \ce{TaTe2}}.
\newblock \emph{\bibinfo{journal}{Phys. Rev. B}} \textbf{\bibinfo{volume}{98}},
  \bibinfo{pages}{224104} (\bibinfo{year}{2018}).

\bibitem{Che18}
\bibinfo{author}{Chen, C.} \emph{et~al.}
\newblock \bibinfo{title}{Trimer bonding states on the surface of the
  transition-metal dichalcogenide \ce{TaTe2}}.
\newblock \emph{\bibinfo{journal}{Phys. Rev. B}} \textbf{\bibinfo{volume}{98}}
  (\bibinfo{year}{2018}).

\bibitem{Bag18}
\bibinfo{author}{El~Baggari, I.}, \bibinfo{author}{Stiehl, G.~M.},
  \bibinfo{author}{Waelder, J.}, \bibinfo{author}{Ralph, D.~C.} \&
  \bibinfo{author}{Kourkoutis, L.~F.}
\newblock \bibinfo{title}{Atomic-resolution cryo-{STEM} imaging of a structural
  phase transition in \ce{TaTe2}}.
\newblock \emph{\bibinfo{journal}{Microsc. Microanal.}}
  \textbf{\bibinfo{volume}{24}}, \bibinfo{pages}{86--87}
  (\bibinfo{year}{2018}).

\bibitem{Wan20}
\bibinfo{author}{Wang, H.} \emph{et~al.}
\newblock \bibinfo{title}{Charge density wave and atomic trimerization in
  layered transition-metal dichalcogenides \ce{1T-MX2}~materials}.
\newblock \emph{\bibinfo{journal}{EPL}} \textbf{\bibinfo{volume}{130}}
  (\bibinfo{year}{2020}).

\bibitem{Svetin_2017}
\bibinfo{author}{Svetin, D.}, \bibinfo{author}{Vaskivskyi, I.},
  \bibinfo{author}{Brazovskii, S.} \& \bibinfo{author}{Mihailovic, D.}
\newblock \bibinfo{title}{Three-dimensional resistivity and switching between
  correlated electronic states in \ce{1T-TaS2}}.
\newblock \emph{\bibinfo{journal}{Sci. Rep.}} \textbf{\bibinfo{volume}{7}}
  (\bibinfo{year}{2017}).

\bibitem{LeBlanc_2010}
\bibinfo{author}{LeBlanc, A.} \& \bibinfo{author}{Nader, A.}
\newblock \bibinfo{title}{Resistivity anisotropy and charge density wave in
  \ce{2H-NbSe2} and \ce{2H-TaSe2}}.
\newblock \emph{\bibinfo{journal}{Solid State Commun.}}
  \textbf{\bibinfo{volume}{150}}, \bibinfo{pages}{1346--1349}
  (\bibinfo{year}{2010}).

\bibitem{Chen_2017}
\bibinfo{author}{Chen, H.}, \bibinfo{author}{Li, Z.}, \bibinfo{author}{Guo, L.}
  \& \bibinfo{author}{Chen, X.}
\newblock \bibinfo{title}{Anisotropic magneto-transport and magnetic properties
  of low-temperature phase of \ce{TaTe2}}.
\newblock \emph{\bibinfo{journal}{{EPL}}} \textbf{\bibinfo{volume}{117}},
  \bibinfo{pages}{27009} (\bibinfo{year}{2017}).

\bibitem{HiRESsims}
\bibinfo{author}{Filippetto, D.} \& \bibinfo{author}{Qian, H.}
\newblock \bibinfo{title}{Design of a high-flux instrument for ultrafast
  electron diffraction and microscopy}.
\newblock \emph{\bibinfo{journal}{J. Phys. B}} \textbf{\bibinfo{volume}{49}},
  \bibinfo{pages}{104003} (\bibinfo{year}{2016}).

\bibitem{ji_ultrafast_2019}
\bibinfo{author}{Ji, F.} \emph{et~al.}
\newblock \bibinfo{title}{Ultrafast relativistic electron nanoprobes}.
\newblock \emph{\bibinfo{journal}{Commun. Phys.}} \textbf{\bibinfo{volume}{2}}
  (\bibinfo{year}{2019}).

\bibitem{Siddiqui_2020}
\bibinfo{author}{Siddiqui, K.} \emph{et~al.}
\newblock \bibinfo{title}{Ultrafast structural dynamics of materials captured
  by relativistic electron bunches}.
\newblock \emph{\bibinfo{journal}{Proc. SPIE 11497, Ultrafast Nonlinear Imaging
  and Spectroscopy {VIII}, 114970J}}  (\bibinfo{year}{2020}).

\bibitem{Vernes_1998}
\bibinfo{author}{Vernes, A.}, \bibinfo{author}{Ebert, H.},
  \bibinfo{author}{Bensch, W.}, \bibinfo{author}{Heid, W.} \&
  \bibinfo{author}{Näther, C.}
\newblock \bibinfo{title}{Crystal structure, electrical properties and
  electronic band structure of tantalum ditelluride}.
\newblock \emph{\bibinfo{journal}{J. Condens. Matter Phys.}}
  \textbf{\bibinfo{volume}{10}}, \bibinfo{pages}{761--774}
  (\bibinfo{year}{1998}).

\bibitem{Erasmus2012}
\bibinfo{author}{Erasmus, N.} \emph{et~al.}
\newblock \bibinfo{title}{Ultrafast dynamics of charge density waves in
  {4H}$_{b}$-\ce{TaSe2} probed by femtosecond electron diffraction}.
\newblock \emph{\bibinfo{journal}{Phys. Rev. Lett.}}
  \textbf{\bibinfo{volume}{109}}, \bibinfo{pages}{167402}
  (\bibinfo{year}{2012}).

\bibitem{Kirkland_2010}
\bibinfo{author}{Kirkland, E.~J.}
\newblock \emph{\bibinfo{title}{Advanced Computing in Electron Microscopy}}
  (\bibinfo{publisher}{Springer {US}}, \bibinfo{year}{2010}).

\bibitem{Ophus_2017}
\bibinfo{author}{Ophus, C.}
\newblock \bibinfo{title}{A fast image simulation algorithm for scanning
  transmission electron microscopy}.
\newblock \emph{\bibinfo{journal}{Advanced Structural and Chemical Imaging}}
  \textbf{\bibinfo{volume}{3}} (\bibinfo{year}{2017}).

\bibitem{Storeck_2020}
\bibinfo{author}{Storeck, G.} \emph{et~al.}
\newblock \bibinfo{title}{Structural dynamics of incommensurate charge-density
  waves tracked by ultrafast low-energy electron diffraction}.
\newblock \emph{\bibinfo{journal}{Struct. Dyn.}} \textbf{\bibinfo{volume}{7}},
  \bibinfo{pages}{034304} (\bibinfo{year}{2020}).

\bibitem{Zewail2008buckling}
\bibinfo{author}{Kwon, O.-H.}, \bibinfo{author}{Barwick, B.},
  \bibinfo{author}{Park, H.~S.}, \bibinfo{author}{Baskin, J.~S.} \&
  \bibinfo{author}{Zewail, A.~H.}
\newblock \bibinfo{title}{Nanoscale mechanical drumming visualized by {4D}
  electron microscopy}.
\newblock \emph{\bibinfo{journal}{Nano Lett.}} \textbf{\bibinfo{volume}{8}},
  \bibinfo{pages}{3557--3562} (\bibinfo{year}{2008}).

\bibitem{Lindenberg2015MoS2Buckling}
\bibinfo{author}{Mannebach, E.~M.} \emph{et~al.}
\newblock \bibinfo{title}{Dynamic structural response and deformations of
  monolayer \ce{MoS2} visualized by femtosecond electron diffraction}.
\newblock \emph{\bibinfo{journal}{Nano Lett.}} \textbf{\bibinfo{volume}{15}},
  \bibinfo{pages}{6889--6895} (\bibinfo{year}{2015}).
  
  \bibitem{Ubaldini2013}
\bibinfo{author}{Ubaldini, A.}, \bibinfo{author}{Jacimovic, J.},
  \bibinfo{author}{Ubrig, N.} \& \bibinfo{author}{Giannini, E.}
\newblock \bibinfo{title}{Chloride-driven chemical vapor transport method for
  crystal growth of transition metal dichalcogenides}.
\newblock \emph{\bibinfo{journal}{Cryst. Growth Des.}}
  \textbf{\bibinfo{volume}{13}}, \bibinfo{pages}{4453--4459}
  (\bibinfo{year}{2013}).

\bibitem{Castellanos_Gomez_2014}
\bibinfo{author}{Castellanos-Gomez, A.} \emph{et~al.}
\newblock \bibinfo{title}{Deterministic transfer of two-dimensional materials
  by all-dry viscoelastic stamping}.
\newblock \emph{\bibinfo{journal}{2D Mater.}} \textbf{\bibinfo{volume}{1}},
  \bibinfo{pages}{011002} (\bibinfo{year}{2014}).

\end{thebibliography}

\newpage

\begin{thebibliography}{99}
\makeatletter
\addtocounter{NAT@ctr}{53}
\makeatother

\bibitem{Blochl1994}
\bibinfo{author}{Bl\"ochl, P.~E.}
\newblock \bibinfo{title}{Projector augmented-wave method}.
\newblock \emph{\bibinfo{journal}{Phys. Rev. B}} \textbf{\bibinfo{volume}{50}},
  \bibinfo{pages}{17953--17979} (\bibinfo{year}{1994}).

\bibitem{Kresse1996}
\bibinfo{author}{Kresse, G.} \& \bibinfo{author}{Furthm\"uller, J.}
\newblock \bibinfo{title}{Efficient iterative schemes for ab initio
  total-energy calculations using a plane-wave basis set}.
\newblock \emph{\bibinfo{journal}{Phys. Rev. B}} \textbf{\bibinfo{volume}{54}},
  \bibinfo{pages}{11169--11186} (\bibinfo{year}{1996}).

\bibitem{Kresse1999}
\bibinfo{author}{Kresse, G.} \& \bibinfo{author}{Joubert, D.}
\newblock \bibinfo{title}{From ultrasoft pseudopotentials to the projector
  augmented-wave method}.
\newblock \emph{\bibinfo{journal}{Phys. Rev. B}} \textbf{\bibinfo{volume}{59}},
  \bibinfo{pages}{1758--1775} (\bibinfo{year}{1999}).

\bibitem{chen2018trimer}
\bibinfo{author}{Chen, C.} \emph{et~al.}
\newblock \bibinfo{title}{Trimer bonding states on the surface of the
  transition-metal dichalcogenide \ce{TaTe2}}.
\newblock \emph{\bibinfo{journal}{Physical Review B}}
  \textbf{\bibinfo{volume}{98}}, \bibinfo{pages}{195423}
  (\bibinfo{year}{2018}).

\bibitem{giustino_electron-phonon_2017}
\bibinfo{author}{Giustino, F.}
\newblock \bibinfo{title}{Electron-phonon interactions from first principles}.
\newblock \emph{\bibinfo{journal}{Rev. Mod. Phys.}}
  \textbf{\bibinfo{volume}{89}}, \bibinfo{pages}{015003}
  (\bibinfo{year}{2017}).

\bibitem{giannozzi_quantum_2009}
\bibinfo{author}{Giannozzi, P.} \emph{et~al.}
\newblock \bibinfo{title}{{QUANTUM} {ESPRESSO}: a modular and open-source
  software project for quantum simulations of materials}.
\newblock \emph{\bibinfo{journal}{J. Phys. Condens. Matter}}
  \textbf{\bibinfo{volume}{21}}, \bibinfo{pages}{395502}
  (\bibinfo{year}{2009}).

\bibitem{wei2017TEMTaSeTe}
\bibinfo{author}{Wei, L.-L.} \emph{et~al.}
\newblock \bibinfo{title}{Charge density wave states and structural transition
  in layered chalcogenide {TaSe}$_{2-x}${Te}$_{x}$}.
\newblock \emph{\bibinfo{journal}{Chin. Phys. Lett.}}
  \textbf{\bibinfo{volume}{34}}, \bibinfo{pages}{086101}
  (\bibinfo{year}{2017}).

\bibitem{wang2020TEMMX2}
\bibinfo{author}{Wang, H.} \emph{et~al.}
\newblock \bibinfo{title}{Charge density wave and atomic trimerization in
  layered transition-metal dichalcogenides \ce{1T-MX2}~materials}.
\newblock \emph{\bibinfo{journal}{EPL}} \textbf{\bibinfo{volume}{130}},
  \bibinfo{pages}{47001} (\bibinfo{year}{2020}).

\bibitem{palmer2015CrystalMaker}
\bibinfo{author}{Palmer, D.~C.}
\newblock \bibinfo{title}{Visualization and analysis of crystal structures
  using crystalmaker software}.
\newblock \emph{\bibinfo{journal}{Z. Kristallogr. Cryst. Mater.}}
  \textbf{\bibinfo{volume}{230}}, \bibinfo{pages}{559--572}
  (\bibinfo{year}{2015}).

\bibitem{kirkland1998advanced}
\bibinfo{author}{Kirkland, E.~J.}
\newblock \emph{\bibinfo{title}{Advanced computing in electron microscopy}},
  vol.~\bibinfo{volume}{12} (\bibinfo{publisher}{Springer},
  \bibinfo{year}{1998}).

\bibitem{ophus2017prism}
\bibinfo{author}{Ophus, C.}
\newblock \bibinfo{title}{A fast image simulation algorithm for scanning
  transmission electron microscopy}.
\newblock \emph{\bibinfo{journal}{Advanced structural and chemical imaging}}
  \textbf{\bibinfo{volume}{3}}, \bibinfo{pages}{1--11} (\bibinfo{year}{2017}).

\bibitem{byrnes2016tmm}
\bibinfo{author}{Byrnes, S.~J.}
\newblock \bibinfo{title}{Multilayer optical calculations}.
\newblock \emph{\bibinfo{journal}{ArXiv}} \bibinfo{pages}{Preprint at
  \url{https://arxiv.org/abs/1603.02720}} (\bibinfo{year}{2016}).

\bibitem{luke2015Si3N4nk}
\bibinfo{author}{Luke, K.}, \bibinfo{author}{Okawachi, Y.},
  \bibinfo{author}{Lamont, M.~R.}, \bibinfo{author}{Gaeta, A.~L.} \&
  \bibinfo{author}{Lipson, M.}
\newblock \bibinfo{title}{Broadband mid-infrared frequency comb generation in a
  \ce{Si3N4} microresonator}.
\newblock \emph{\bibinfo{journal}{Opt. Lett.}} \textbf{\bibinfo{volume}{40}},
  \bibinfo{pages}{4823--4826} (\bibinfo{year}{2015}).

\bibitem{schinke2015SiRefractiveIndex}
\bibinfo{author}{Schinke, C.} \emph{et~al.}
\newblock \bibinfo{title}{Uncertainty analysis for the coefficient of
  band-to-band absorption of crystalline silicon}.
\newblock \emph{\bibinfo{journal}{AIP Adv.}} \textbf{\bibinfo{volume}{5}},
  \bibinfo{pages}{067168} (\bibinfo{year}{2015}).

\bibitem{rodriguez2016SiO2nk}
\bibinfo{author}{Rodr{\'\i}guez-de Marcos, L.~V.}, \bibinfo{author}{Larruquert,
  J.~I.}, \bibinfo{author}{M{\'e}ndez, J.~A.} \& \bibinfo{author}{Azn{\'a}rez,
  J.~A.}
\newblock \bibinfo{title}{Self-consistent optical constants of \ce{SiO2} and
  \ce{Ta2O5} films}.
\newblock \emph{\bibinfo{journal}{Opt. Mater. Express}}
  \textbf{\bibinfo{volume}{6}}, \bibinfo{pages}{3622--3637}
  (\bibinfo{year}{2016}).

\bibitem{durhamScopeFoundry}
\bibinfo{author}{Durham, D.~B.}, \bibinfo{author}{Ogletree, D.~F.} \&
  \bibinfo{author}{Barnard, E.~S.}
\newblock \bibinfo{title}{Scanning {Auger} spectromicroscopy using the
  {ScopeFoundry} software platform}.
\newblock \emph{\bibinfo{journal}{Surf. Interface Anal.}}
  \textbf{\bibinfo{volume}{50}}, \bibinfo{pages}{1174--1179}
  (\bibinfo{year}{2018}).

\bibitem{ScopeFoundryWebsite}
\bibinfo{author}{Barnard, E.~S.}
\newblock \bibinfo{title}{{ScopeFoundry}: A python platform for controlling
  custom laboratory experiments and visualizing scientific data}.
\newblock \bibinfo{howpublished}{\url{http://www.scopefoundry.org/}}.

\end{thebibliography}
\end{document}